\documentclass[journal,comsoc]{IEEEtran}

\usepackage{epsfig}
\usepackage[english]{babel}
\usepackage[utf8x]{inputenc}
\usepackage[T1]{fontenc}

\usepackage{cite}
\usepackage{textcomp}
\usepackage{xcolor}

\usepackage{amsmath}
\usepackage[cmintegrals]{newtxmath}

\usepackage{amsfonts}
\usepackage{mathrsfs}
\usepackage{graphicx}
\usepackage{booktabs}
\usepackage[labelfont=bf]{caption}
\usepackage{subcaption}
\usepackage{verbatim} 
\usepackage{float} 
\usepackage[algoruled,noline,longend,linesnumbered]{algorithm2e}
\DeclareMathOperator*{\E}{\mathbb{E}}

\DeclareMathOperator*{\Sspace}{\mathcal{S}}
\DeclareMathOperator*{\A}{\mathcal{A}}
\DeclareMathOperator*{\Nagents}{\mathcal{N}}

\def\BibTeX{{\rm B\kern-.05em{\sc i\kern-.025em b}\kern-.08em
    T\kern-.1667em\lower.7ex\hbox{E}\kern-.125emX}}

\usepackage[colorlinks = true,
            linkcolor = black,
            urlcolor  = black,
            citecolor = black,
            anchorcolor = blue]{hyperref}    

\graphicspath{{./}{../}}

\usepackage{tikz}
\usetikzlibrary{calc} 

\begin{document}

%

\IEEEpubid{\begin{minipage}{\textwidth} \vspace*{1cm} \centering This is the author's version of DOI 10.1109/TCCN.2023.3235719 published in IEEE Transactions on Cognitive Communications and Networking. \\ 
\copyright~2023 IEEE. Personal use is permitted, but republication/redistribution requires IEEE permission. \end{minipage}}

\title{MAGNNETO: A Graph Neural Network-based Multi-Agent system for Traffic Engineering}
\label{title}

\author{Guillermo Bernárdez, José Suárez-Varela, Albert López, Xiang Shi, Shihan Xiao, Xiangle Cheng, Pere~Barlet-Ros, and Albert~Cabellos-Aparicio
\thanks{G. Bernárdez, J. Suárez-Varela, A. López, P. Barlet-Ros and A. Cabellos-Aparicio are with Barcelona Neural Networking Center, Universitat Politècnica de Catalunya, Barcelona, Spain. Contact: guillermo.bernardez@upc.edu}
\thanks{X.~Shi, S. Xiao and X. Cheng are with the Network Technology Lab., Huawei Technologies Co., Ltd., Beijing, China.}
}

\maketitle

\begin{abstract}
\addcontentsline{toc}{section}{Abstract}
Current trends in networking propose the use of Machine Learning (ML) for a wide variety of network optimization tasks. As such, many efforts have been made to produce ML-based solutions for Traffic Engineering (TE), which is a fundamental problem in ISP networks. Nowadays, state-of-the-art TE optimizers rely on traditional optimization techniques, such as Local search, Constraint Programming, or Linear programming. In this paper, we present MAGNNETO, a distributed ML-based framework that leverages Multi-Agent Reinforcement Learning and Graph Neural Networks for distributed TE optimization. MAGNNETO deploys a set of agents across the network that learn and communicate in a distributed fashion via message exchanges between neighboring agents. Particularly, we apply this framework to optimize link weights in OSPF, with the goal of minimizing network congestion. In our evaluation, we compare MAGNNETO against several state-of-the-art TE optimizers in more than 75 topologies (up to 153 nodes and 354 links), including realistic traffic loads. Our experimental results show that, thanks to its distributed nature, MAGNNETO achieves comparable performance to state-of-the-art TE optimizers with significantly lower execution times. Moreover, our ML-based solution demonstrates a strong generalization capability to successfully operate in new networks unseen during training. 
\end{abstract}


\section{Introduction}\label{sec:introduction}

During the last decade, the networking community has devoted significant efforts to build efficient solutions for automated network control, pursuing the ultimate goal of achieving the long-desired \textit{self-driving networks}~\cite{feamster2017and, mestres2017knowledge}. In this vein, Machine Learning (ML) is considered as a promising technique for producing efficient tools for autonomous networking~\mbox{\cite{wang2017machine,boutaba2018comprehensive}}.

In this paper, we revisit a fundamental networking problem: Traffic Engineering (TE) optimization~\cite{RFC2702,RFC3272}. TE is among the most common operation tasks in today's ISP networks. Here, the classical optimization goal is to minimize network congestion, which is typically achieved by minimizing the maximum link utilization in the network~\cite{azar2004optimal, fortz2004increasing, gay2017expect, hartert2015declarative, fortz2000internet}. Given the relevance of this problem, we have witnessed a plethora of proposals approaching this problem from different angles, such as optimizing the configuration of widely deployed link-state protocols (e.g., OSPF~\cite{RFC2328}), making fine-grained flow-based routing, or re-routing traffic across overlay networks~\cite{wang2008overview,mendiola2016survey}. 

Likewise, for the last years the networking community has focused on developing effective ML-based solutions for TE. In particular, many works propose the use of Reinforcement Learning (RL) for efficient TE optimization (e.g.,~\mbox{\cite{valadarsky2017learning,xu2018experience,drl-giorgio,geng2020multi}}). However, at the time of this writing, no ML-based proposal has succeeded to replace long-established TE solutions; indeed, the best performing TE optimizers to date are based on traditional optimization algorithms, such as Constraint Programming~\cite{hartert2015declarative}, Local Search~\cite{gay2017expect}, or Linear Programming~\mbox{\cite{bhatia2015optimized,fortz2000internet}}.

In this paper, we present MAGNNETO, a novel ML framework for distributed TE optimization leveraging Graph Neural Networks~(GNN)~\cite{scarselli2008graph} and Multi-Agent Reinforcement Learning (MARL)~\cite{foerster2018deep} at its core\footnote{MAGNNETO stands for \underline{M}ulti-\underline{A}gent \underline{G}raph \underline{N}eural \underline{Net}work \underline{O}ptimization. The code of this framework and all the data needed to reproduce our experiments are publicly available at: \url{https://github.com/BNN-UPC/Papers/wiki/MAGNNETO-TE}.}. In the proposed algorithm, a RL-based agent is deployed on each router. Similarly to standard intradomain routing protocols (e.g., OSPF), \mbox{MAGNNETO's} agents exchange information with their neighbors in a distributed manner. In particular, agents communicate via a neural network-driven message passing mechanism, and learn how to cooperate to pursue a common optimization goal. As a result, the proposed framework is fully distributed, and agents learn how to effectively communicate to perform intradomain TE optimization, i.e. to minimize the maximum link utilization in the network.

\IEEEpubidadjcol

More in detail, MAGNNETO presents the following contributions:

\textbf{Top performance with very low execution times:}
We compare MAGNNETO against a curated set of well-established TE solutions: SRLS~\cite{gay2017expect}, DEFO~\cite{hartert2015declarative} and \mbox{TabuIGPWO~\cite{fortz2000internet}}. These solutions implement mature optimization techniques on top of expert knowledge. As a result, they are able to achieve close-to-optimal performance in large-scale networks within minutes~\cite{gay2017repetita}. Our results show that MAGNNETO achieves comparable performance to these state-of-the-art TE solutions, while being significantly faster. In fact, when enabling several simultaneous actions in our framework, it runs up to three orders of magnitude faster than the baseline optimizers (sub-second vs. minutes) in networks with $100+$ nodes. The reason for this is the fully decentralized architecture of MAGNNETO, which naturally distributes and parallelizes the execution across the network. 
\vspace{0.1cm}

\textbf{Generalization over unseen networks:}
A common downside of current ML-based solutions applied to networking is their limited performance when operating in different networks to those seen during training, which is commonly referred to as lack of \emph{generalization}~\cite{battaglia2018relational}. Without generalization, training \emph{must} be done at the same network where the ML-based solution is expected to operate. Hence, from a practical standpoint generalization is a crucial aspect, as training directly in networks in production is typically unfeasible. MAGNNETO implements internally a GNN, which introduces proper learning biases to generalize across networks of different sizes and structures~\cite{battaglia2018relational}. In our evaluation, we train MAGNNETO in two different networks, and test its performance and speed on 75 real-world topologies from the Internet Topology Zoo~\cite{knight2011internet} not seen before. Our results show that in such scenarios, MAGNNETO still achieves comparable performance to state-of-the-art TE optimizers, while being significantly faster.

\textbf{No need for overlay technologies}: 
Recent TE optimizers rely on novel overlay technologies to achieve their optimization goals~\cite{gay2017expect,hartert2015declarative}. By leveraging Segment Routing~\cite{filsfils2015segment} these solutions are able to use arbitrary overlay paths that are not routed via the standard OSPF weights. This allows to extend the routing space to a source-destination granularity and --as shown in the literature-- it renders outstanding results. However, in this paper we show that comparable performance is achievable by using only standard destination-based OSPF routing. Indeed, MAGNNETO is fully compliant with current OSPF-based networks, and does not require the use of any overlay technology.
\vspace{0.1cm}

MAGNNETO is partially based on an earlier version presented at~\cite{icnp-paper}. In that work, we raised an open question: \emph{\mbox{Is ML ready} for Traffic Engineering optimization?}
Our goal was to discuss whether state-of-the-art ML techniques are mature enough to outperform traditional TE solutions; to this end, we presented a ML framework for TE optimization, and made an exploratory evaluation on this. This paper actually deeps dive into this question by formulating an enhanced ML framework --MAGNNETO-- and performing a much more comprehensive evaluation. We summarize below the main novelties of this work with respect to \cite{icnp-paper}:
\begin{itemize}
\item MAGNNETO formulates the TE problem as a Decentralized Partially-Observable Markov Decision Process (Dec-POMDP), which enables to achieve a more functional MARL setting. Instead, the previous solution~\cite{icnp-paper} operated over a classical MDP, where agents must take actions sequentially in a synchronized manner.

\item  MAGNNETO supports simultaneous actions at each RL optimization step. This dramatically reduces the execution time (up to 10x in our experiments) with respect to the previous framework, which was limited by design to one action per step.

\item We present in this paper an extensive evaluation including 75+ real-world topologies, large-scale scenarios (up to 153 nodes), and a benchmark consisting of a representative collection of advanced TE optimizers. In contrast, the evaluation of \cite{icnp-paper} only considered 3 different topologies of limited size (up to 24 nodes), and the results were compared against a single TE solver.
\end{itemize}

The remainder of this paper is as follows. Section~\ref{sec:scenario} describes the TE scenario where we deploy the proposed ML-based system. Section~\ref{sec:architecture} formalizes \mbox{MAGNNETO}, as a general framework for networked environments. Afterwards, Section~\ref{sec:routing} describes how we adapt this framework to perform intradomain TE optimization. In Section~\ref{sec:results}, we make an extensive evaluation of the proposed framework against state-of-the-art TE proposals. Section~\ref{sec:related} summarizes the main existing works related to this paper, and lastly Section~\ref{sec:conclusions} concludes the paper.

\section{Network Scenario} \label{sec:scenario}

\begin{figure}[!t]
\centering
    \includegraphics[width=1.0\columnwidth]{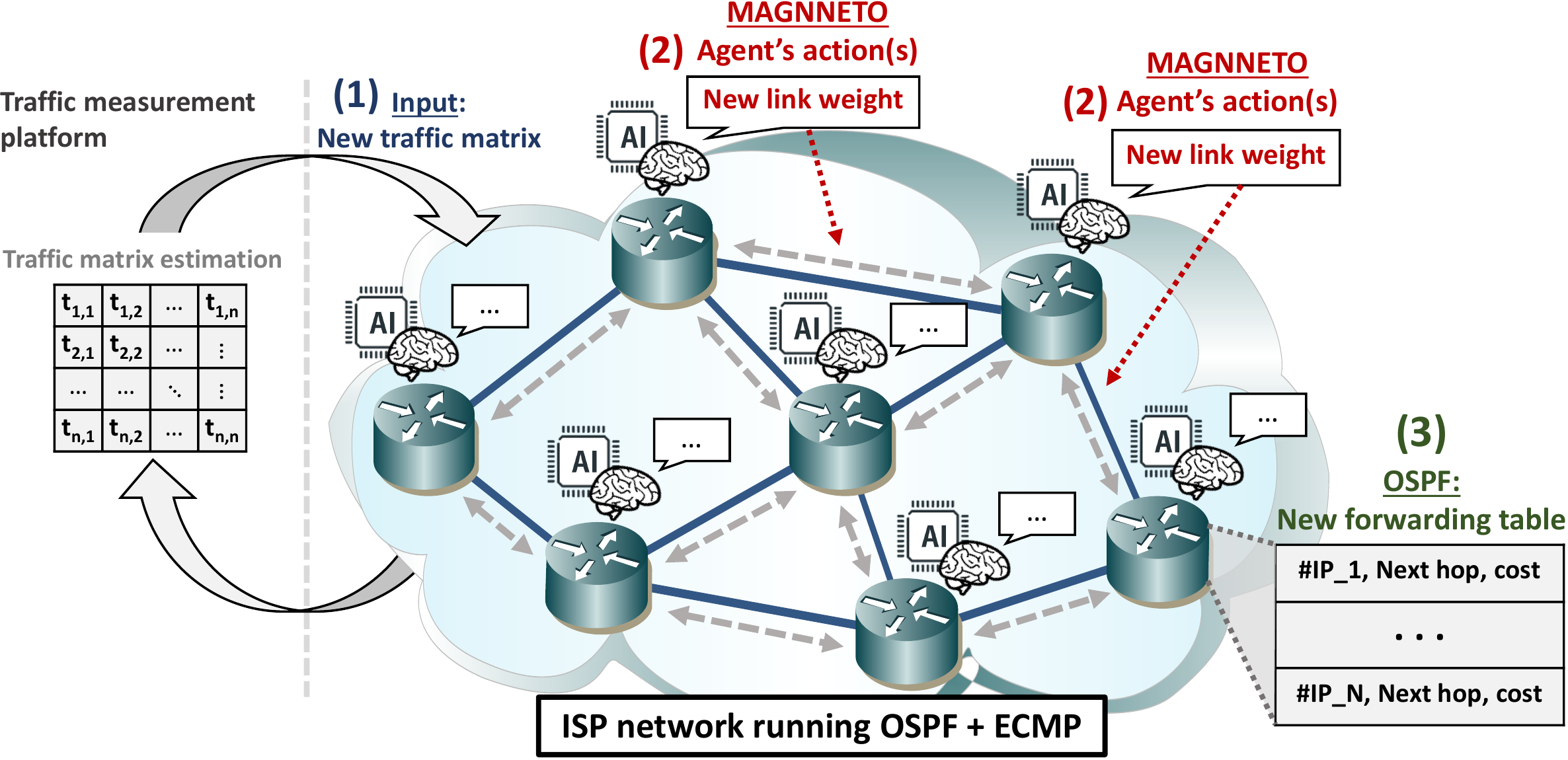}
  \caption{Intradomain traffic engineering optimization with MAGNNETO.}
  \label{fig:scenario}
\end{figure}

This section describes the intradomain TE scenario where MAGNNETO operates. In this paper, we consider the intradomain TE problem, where network traffic is measured and routed to minimize network congestion. Typically, IP networks run link-state Interior Gateway Protocols (IGP), such as Open Shortest Path First (OSPF)~\cite{RFC2328}, that choose paths using the Dijkstra's algorithm over some \mbox{pre-defined} link weights. 

There exists a wide range of architectures and algorithms for TE in the literature~\cite{huang2018survey}. 
Network operators commonly use commercial tools~\cite{cisco2013mate,juniper} to fine-tune link weights. However, other mechanisms propose to add extra routing entries~\cite{sridharan2005achieving} or end-to-end tunnels (e.g., RSVP-TE~\cite{minei2005mpls}) to perform source-destination routing, thus expanding the solution space. 

MAGNNETO is a fully distributed framework that interfaces with standard OSPF, by optimizing the link weights used by such protocol. As a result, it does not require any changes to OSPF and it can be implemented with a software update on the routers where it is deployed. In this context, relying on well-known link-state routing protocols, such as OSPF, offers the advantage that the network is easier to manage compared to finer-grained alternatives, such as flow-based routing~\cite{xu2011link}. 

Figure \ref{fig:scenario} illustrates the general operational workflow of MAGNNETO:

\textbf{1) Traffic Measurement:} First, a traffic measurement platform deployed over the network identifies a new Traffic Matrix (TM). This new TM is communicated to all participating routers (Fig.~\ref{fig:scenario}, step 1), which upon reception will start the next step and optimize the routing for this TM. We leave out of the scope of this paper the details of this process, as TM estimation is an extensive research field with many established proposals. For instance, this process can be done periodically (e.g., each 5-10 minutes as in \cite{fortz2000internet}), where the TM is first estimated and then optimized. Some proposals trigger the optimization process when a relevant change is detected in the TM \cite{benson2011microte}, while others use prediction techniques to optimize it in advance \cite{luo2013dsox}. Also, some real-world operators make estimates considering their customers' subscriptions and operate based on a static TM. Our proposal is flexible and can operate with any of these approaches. 

\textbf{2) MAGNNETO TE optimization:} Once routers receive the new TM, the distributed RL-based agents of MAGNNETO start the TE optimization process, which eventually computes the per-link weights that optimize OSPF routing in the subsequent step (Fig.~\ref{fig:scenario}, step 2). Particularly, we set the goal to minimize the maximum link load (\mbox{\emph{MinMaxLoad}}), which is a classic TE goal in carrier-grade networks~\cite{azar2004optimal, fortz2004increasing,hartert2015declarative}. This problem is known to be NP-hard, and even good settings of the weights can deviate significantly from the optimal configuration~\cite{xu2011link,fortz2004increasing}. Our MARL optimization system is built using a distributed Graph Neural Network (GNN) that exchanges messages over the physical network topology. Messages are sent between routers and their directly attached neighbors. The content of such messages are \emph{hidden states} that are produced and consumed by artificial neural networks and do not have a human-understandable \emph{meaning}. The GNN makes several message iterations and, during this phase, local configuration of the router remains unchanged, thus having no impact on the current traffic. More details about the inner workings, performance, communication overhead, and computational cost can be found in Sections~\ref{sec:architecture}-\ref{sec:results}.

\textbf{3) OSPF convergence:} Lastly, the standard OSPF convergence process is executed taking into account the new per-link weights computed by MAGNNETO. Specifically, each agent has computed the optimal weigths for its locally attached links. For OSPF to recompute the new forwarding tables, it needs to broadcast the new link weights; this is done using the standard OSPF Link-State Advertisements (LSAs) \cite{RFC2328}. Once the routers have an identical view of the network, they compute locally their new forwarding tables (Fig.~\ref{fig:scenario}, step 3), and traffic is routed following the optimization goal. Convergence time of OSPF is a well-studied subject. For instance, routing tables can converge in the order of a few seconds in networks with thousands of links \cite{basu2001stability}.

\section{MAGNNETO} \label{sec:architecture}
This section provides a detailed description on how MAGNNETO operates. To do so we first briefly introduce the main ML methodologies it implements. Note that MAGNNETO is conceived as a general framework to optimize networked environments in a distributed fashion; details on how it is particularly adapted to face intradomain TE are then provided in Section~\ref{sec:routing}.

\subsection{Related ML-based Technologies} \label{sec:back}

MAGNNETO incorporates two well-known ML-based mechanisms: Multi-Agent Reinforcement Learning and Graph Neural Networks. Let us provide some background on these technologies:

\subsubsection{Reinforcement Learning (RL)} \label{subsec:back-RL}

According to the regular setting of RL~\cite{bertsekas1996neuro}, an agent interacts with the environment in the following way: at each step $t$, the agent selects an action $a_t$ based on its current state $s_t$, to which the environment responds with a reward $r_t$ and then moves to the next state $s_{t+1}$. This interaction is modeled as an episodic, time-homogeneous Markov Decision Process (MDP) $(\Sspace, \A, r, P, \gamma)$, where $\Sspace$ and $\A$ are respectively the state and action spaces; $P$ is the transition kernel, $s_{t+1} \sim P(\cdot | s_t,a_t)$; $r_t$ represents the immediate reward given by the environment after taking action $a_t$ from state $s_t$; and $\gamma \in (0,1]$ is the discount factor used to compute the return $G_t$, defined as the --discounted-- cumulative reward from a certain time-step $t$ to the end of the episode $T$: $G_t = \sum_{t=0}^T \gamma^t r_t$. The behavior of the agent is described by a policy $\pi: \mathcal{S} \to \mathcal{A}$, which maps each state to a probability distribution over the action space, and the goal of an RL agent is to find the optimal policy in the sense that, given any considered state $s\in \Sspace$, it always selects an action that maximizes the expected return~$\hat{G}_t$.
There are two main model-free approaches to this end~\cite{sutton2018reinforcement}: 
\begin{itemize}
    \item Action-value methods, typically referred to as \mbox{Q-learning}; the policy $\pi$ is indirectly defined from the learned estimates of the action value function 
    $Q^\pi (s,a)~=~\E_{\pi} \left[ G_t | s_0=s, a_0=a \right]$.
    \item Policy Gradient (PG) methods, which directly attempt to learn a parameterized policy representation $\pi_\theta$. The Actor-Critic family of PG algorithms also involves learning a function approximator $V_\phi(s)$ of the state value function
    $V^{\pi_\theta} (s) = \E _{\pi_\theta} \left[ G_t | s_t=s \right]$. In this case, actions are exclusively selected from function $\pi_\theta$, which estimates the policy (i.e., the actor), but the training of such policy is guided by the estimated value function $V_\phi(s)$, which assesses the consequences of the actions taken (i.e., the critic). 
\end{itemize}

\subsubsection{Multi-Agent Reinforcement Learning (MARL)} \label{subsec:back-MARL}

In a MARL framework there is a set of agents $\mathcal{V}$ interacting with a common environment that have to learn how to cooperate to pursue a common goal. Such a setting is generally formulated as a Decentralized Partially Observable MDP (Dec-POMDP)~\cite{foerster2018deep} where, besides the global state space $\Sspace$ and action space $\A$, it distinguishes local state and action spaces for every agent --i.e., $\Sspace_v$ and $\A_v$ for $v \in \mathcal{V}$. At each time step~$t$ of an episode, each agent may choose an action $a^v_t \in \A_v$ based on local observations of the environment encoded in its current state $s^v_t \in \Sspace_v$. Then, the environment produces individual rewards $r_t^v$ (and/or a global reward $r_t$), and it evolves to a next global state $s_{t+1}\in \Sspace$ --i.e., each agent~$v$ transitions to the following state $s^v_{t+1} \in \Sspace_v$. Typically, a MARL system seeks for the optimal global policy by learning a set of local policies $\{\pi_{\theta_v}\}_{v \in \mathcal{V}}$. For doing so, most state-of-the-art MARL solutions implement traditional (single-agent) RL algorithms on each distributed agent, while incorporating some kind of cooperation mechanism between them~\cite{foerster2018deep}. The standard approach for obtaining a robust decentralized execution, however, is based on a centralized training where extra information can be used to guide agents' learning~\cite{oliehoek2008}.

\subsubsection{Graph Neural Networks (GNN)} \label{subsec:back-GNN}

 These models are a recent family of neural networks specifically conceived to operate over graph-structured data~\cite{scarselli2008graph, battaglia2018relational}. Among the numerous GNN variants developed to date~\cite{wu2020comprehensive}, 
we focus on Message Passing Neural Networks (MPNN)~\cite{gilmer2017neural}, which is a well-known type of GNN whose operation is based on an iterative message-passing algorithm that propagates information 
between elements in a graph \mbox{$\mathcal{G} = (\Nagents, \mathcal{E})$}. Focusing on the set of nodes, the process is as follows: 
first, each node $v \in \Nagents$ initializes its hidden state $h_v^0$ using some initial features already included in the input graph. At every message-passing step $k$, each node $v$ receives via messages the current hidden state of all the nodes in its neighborhood $\mathcal{B}(v)=\{u \in \Nagents | \exists e \in \mathcal{E}, e = (u,v) \lor e=(v,u) \}$, and processes them individually by applying a message function \textit{m(·)} together with its own internal state $h_v^k$. Then, the processed messages are combined by an aggregation function \textit{a(·)}:
\begin{equation}
M_v^k = a( \{ m(h_{v}^k, h_{i}^{k}) \}_{i \in \mathcal{B}(v)})
\label{eq:message_function}
\end{equation}
Finally, an update function \textit{u(·)} is applied to each node $v$; taking as input the aggregated messages $M_{v}^{k}$ and its current hidden state $h_v^k$, it outputs a new hidden state for the next step ($k+1$):
\begin{equation}
h_{v}^{k+1} = u(h_{v}^{k}, M_{v}^{k}).
\vspace{0.15cm}
\label{eq:update_function}
\end{equation}
After a certain number of message passing steps $K$, a readout function \textit{r(·)} takes as input the final node states $h_v^K$ to produce the final output of the GNN model. This readout function can predict either features of individual elements (e.g., a node's class) or global properties of the graph.
Note that a MPNN model generates \textit{a single set of message, aggregation, update, and readout functions that are replicated at each selected graph element}.

\subsection{Execution Framework} \label{subsec:framework}

MAGNNETO internally models a networked environment as a graph \mbox{$\mathcal{G} = (\Nagents, \mathcal{E}, \mathcal{V})$}, with $\Nagents$ and $\mathcal{E}$ representing the set of nodes and edges, respectively, and $\mathcal{V}$ acting for a set of agents that can control some of the graph entities (nodes or edges). 
Let $\Sspace$ and $\A$ represent the global state and action spaces, respectively, defined as the joint and union of the respective agents' local spaces, $\Sspace = \prod_{v\in \mathcal{V}} \Sspace_v$ and $\A = \bigcup_{v\in \mathcal{V}} \A_v$.
The theoretical framework of MAGNNETO allows to implement both Q-learning and PG methods, so for the sake of generalization let $f_\theta$ represent the global RL-based function that is aimed to learn --i.e., the global state-action value function $Q_\theta$ for the former, or the global policy $\pi_\theta$ for the latter. 

A main contribution of MAGNNETO is that it makes all agents $v \in \mathcal{V}$ learn the global RL-based function approximator in a fully distributed fashion --i.e., all agents end up constructing and having access to the very same representation $f_\theta$. In particular, and from a theoretical RL standpoint, this allows to formulate the problem within two different paradigms depending on the number of actions allowed at each time-step of the RL episode. On the one hand, imposing a single action per time-step enables to devise the problem as a time-homogeneous MDP of single-agent RL~\cite{sutton2018reinforcement}. On the other hand, it requires the more challenging Dec-POMDP formalization of standard MARL~\cite{foerster2018deep} when letting several agents act simultaneously. Note, however, that in practice the execution pipeline of MAGNNETO is exactly the same in both cases.

Another relevant feature of our design is that all agents \mbox{$v\in \mathcal{V}$} are able to internally construct such global representation $f_\theta$ mainly through message communications with their direct neighboring agents $\mathcal{B}(v)$ and their local computations, no longer needing a centralized entity responsible for collecting and processing all the global information together. Such a decentralized, message-based generation of the global function is achieved by modeling the global function $f_\theta$ with a MPNN (see Sec.~\ref{subsec:back-GNN}), so that all agents $v \in \mathcal{V}$ deployed in the network are actually \textit{replicas} of the MPNN modules (message, aggregation, update and readout functions) that perform regular message exchanges with their neighbors $\mathcal{B}(v)$ following the message passing iteration procedure of MPNNs; in particular, note that such \textit{parameter sharing} implies that all agents share as well the same local state and action spaces. 
This reinterpretation of a MPNN as a set of copies of its internal modules is especially important due to the fact that in our approach we directly map the graph $\mathcal{G}$ to a real networked scenario, deploying copies of the MPNN modules along hardware devices in the network (e.g., routers) and making all message communications involved to actually go through the real network infrastructure. Hence, our proposed architecture naturally distributes the execution of the MPNN, and consequently is able to fully decentralize the execution of single-agent RL algorithms. 

\setlength{\algomargin}{1.5em}
\SetAlCapHSkip{0em}
\begin{algorithm}[!t]
\caption{MAGNNETO's execution pipeline.} \label{alg:pipeline}
\DontPrintSemicolon
\SetKwComment{CustomComment}{\#}{}
\SetKwInput{KwIn}{\hspace{-1.5em} Input}
\SetKwInput{KwOut}{\hspace{-1.5em} Output}
\SetKwInOut{Require}{\hspace{-1.5em} Require}
\SetKw{KwBy}{by}
\Require{A graph $\mathcal{G} = (\Nagents, \mathcal{E})$ with a set of agents $\mathcal{V}$, MPNN trained parameters $\theta=\{\theta_i\}_{i\in \{m,a,u,r\}}$}
\KwIn{Initial graph configuration $X^0_{\mathcal{G}}$, episode length $T$, number of message passing steps $K$}
    \textnormal{Agents initialize their states $s_v^0$ based on $X^0_{\mathcal{G}}$}\;
    \For{$t \leftarrow 0$ \KwTo $T$}{
        \textnormal{Agents initialize their hidden states }$h_v^0 \leftarrow (s_v^t, 0, \dots , 0)$\;
        \For{$k \leftarrow 0$ \KwTo $K$}{
            \textnormal{Agents share their current hidden state $h_v^k$ to neighboring agents $\mathcal{B}(v)$}\;
            \textnormal{Agents process the received messages $M_v^{k} \leftarrow a_{\theta_a}( \{ m_{\theta_m}(h_{v}^k, h_{\mu}^k) \}_{\mu \in \mathcal{B}(v)})$}\;
            \textnormal{Agents update their hidden state $h_{v}^{k+1}\leftarrow u(h_{v}^{k}, M_v^{k})$}\;
        }
        \textnormal{Agents partially evaluate the RL function $f_\theta$ over their own actions $\{  f_\theta(s_t,a) \}_{a \in \A_v} \leftarrow r_{\theta_r}(h_v^K)$}\;
        \textnormal{Agents receive the partial evaluations of $f_\theta$ of the rest of agents and build the global representation $f_\theta \leftarrow \{f_\theta(s_t,a) \}_{a \in \A}$}\; 
        \textnormal{Agents select the same set of actions $A_t$ according to $f_\theta$}\;
        \textnormal{Agents whose action was selected execute it, and the environment updates the graph configuration $X^{t+1}_{\mathcal{G}}$}\;
        \textnormal{Agents update their states $s_v^{t+1}$ based on $X^{t+1}_{\mathcal{G}}$}\;
    }
\KwOut{New graph configuration $X^{*}_{\mathcal{G}}$ that optimizes some pre-defined objective or metric}
\end{algorithm}

Algorithm \ref{alg:pipeline} summarizes the resulting distributed pipeline. 
At each time-step $t$ of an episode of length $T$, the MPNN-driven process of approximating the function $f_\theta(s_t,a_t)$ \mbox{--where} $s_t \in \Sspace$ and $a_t\in \A$ refer to the global state and action at $t$-- first constructs a meaningful hidden state $h_v$ for each agent $v\in\mathcal{V}$. Each hidden state $h_v$ basically depends on the hidden representations of the neighboring agents $\mathcal{B}(v)$, and its initialization $h_v^0$ is a function of the current agent state $s^t_v\in \Sspace_v$, which is in turn based on some pre-defined internal agent features $x_v^t$. 
Those representations are shaped during $K$ message-passing steps, where hidden states are iteratively propagated through the graph via messages between direct neighbors. In particular, successive hidden states $h_v^k$, where $k$ accounts for the message-passing step, are computed by the message, aggregation and update functions of the MPNN, as previously described in Section \ref{subsec:back-GNN}. 

Once agents generate their final hidden representation, a readout function --following the MPNN nomenclature-- is applied to each agent to finally obtain the global function~$f_\theta$. Particularly, in our system the readout is divided into two steps: first, each agent $v\in\mathcal{V}$ implements a local readout that takes as input the final representation $h^K_v$, and outputs the final value -or a representation- of the global function $f_\theta$ over every possible action in the agent's space $\A_v$; for instance, this output could be the unnormalized log probability (i.e., logit) of the agent's actions in case of PG methods, or directly the q-value associated to each action when considering Q-learning algorithms. The second and last steps involve a communication layer that propagates such individual outputs to the rest of the agents, so that all of them can internally construct the global representation of $f_{\theta}$ for the overall network state \mbox{$s_t = \prod_{v\in\mathcal{V}}s^t_v$} and all possible actions $\bigcup_{v\in \mathcal{V}} \{a_{v,0},a_{v,1},\dots, a_{v,i}\}$, with $i\in\mathbb{N} \backslash \{0\}$ the number of actions of local agent spaces $\A_v$. Finally, to ensure that all distributed agents sample the same actions when $f_\theta$ encodes a distribution, they are provided with the same probabilistic seed before initiating the process. Consequently, only agents whose action has been selected does execute an action at each time-step $t$. Note that actions are not actually applied over the network configuration until the whole optimization process finishes.

\section{MAGNNETO for Traffic Engineering} \label{sec:routing}

In this section we describe the particular adaptations of the general MAGNNETO framework when applying it to the intradomain TE scenario described in Section \ref{sec:scenario}. Moreover, we provide some details about the training pipeline of our models.

\begin{figure*}[!t]
\centering
    \includegraphics[width=2.05\columnwidth]{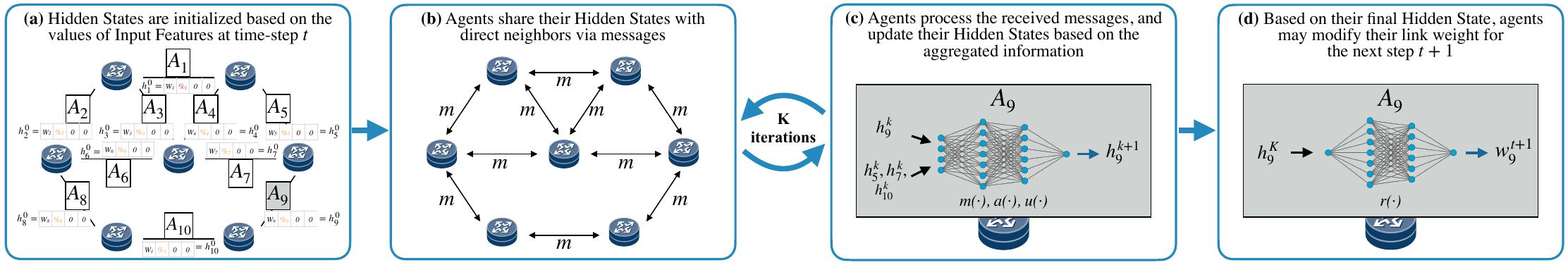}
  \caption{Description of the message passing and action selection process of MAGNNETO at a certain time-step $t$ of an episode. For simplicity, visual representations of steps (c) and (d) are focused on a single agent ($A_{9}$); however, note that the same procedure is executed in parallel in all link-based agents.}
  \label{fig:message_passing}
  \vspace*{-0.2cm}
\end{figure*}

\subsection{General Setting}

A straightforward approach to map the graph $\mathcal{G}$ of the described MAGNNETO framework to a computer network infrastructure is to associate the nodes $\mathcal{N}$ to hardware devices (e.g., router, switches) and the edges $\mathcal{E}$ to the physical links of the network. 
Regarding the set of agents $\mathcal{V}$, they can be identified either with the set of nodes, so that they individually control a hardware device, or with the set of edges by controlling some configuration parameters of a link connecting two devices.

In the intradomain TE problem, the goal is to learn the set of link weights $\mathcal{W} = \{w_e\}_{e \in \mathcal{E}}$ that minimizes the maximum link utilization for a certain traffic matrix $TM$. Hence, we adapt MAGNNETO so that each agent controls a link (i.e., $\mathcal{V}  \widehat{=} \mathcal{E}$) and can modify its weight $w_e$; in fact, in order to make the notation simpler, from now on we will refer to each agent $v \in \mathcal{V}$ as the edge $e \in \mathcal{E}$ it represents. We also note that:
\begin{itemize}
    \item computer networks are commonly represented as directed graphs with links in both directions, so for each directed link \mbox{$e=(n_e^{src},n_e^{dst})\in\mathcal{E}, \textnormal{ with } n_e^{src},n_e^{dst} \in \mathcal{N},$} we define its neighbor as the set $\mathcal{B}(e)$ of edges whose source node coincides with the destination node of $e$, i.e. $\mathcal{B}(e) = \{ e' \in \mathcal{E} | n_{e'}^{src} = n_e^{dst} \}$. In other words, edges in $\mathcal{B}(e)$ are those links that can potentially receive traffic from link $e$.
    \item in practice, link-based agents $e\in \mathcal{E}$ would be deployed and executed in their adjacent source ($n_e^{src}$) or destination ($n_e^{dst}$) hardware device.
\end{itemize}
Furthermore, we implement a well-known Actor-Critic method named Proximal Policy Optimization (PPO)~\cite{schulman2017proximal}, which offers a favorable balance between reliability, sample complexity, and simplicity. Consequently, in this case the global function $f_\theta$ of the framework (see Sec.~\ref{subsec:framework}) is the global policy $\pi_\theta$ of the actor. Regarding the critic's design, more information can be found in Section~\ref{subsec:Training}.

\subsection{Adapting MAGNNETO to TE}

Having clear the general configuration of our MAGNNETO implementation, now we will further describe its operation when dealing with the intradomain TE objective.
To do so, let us reinterpret each of the main fundamental elements introduced earlier from a TE perspective:

\subsubsection{Environment} 
We consider episodes of a fixed number of time-steps $T$. 
At the beginning of each episode, the environment provides with a set of traffic demands between all source-destination pairs (i.e., an estimated traffic matrix~\cite{fortz2000internet}). Each link $e \in \mathcal{E}$ has an associated capacity $c_e$, and it is initialized with a certain link weight $w_e^0$. These link weights are in turn used to compute the routers' forwarding tables, using the standard Dijkstra's algorithm. Each agent $v_e \in \mathcal{V}$ has access to its associated link features, which in our case are the current weight, its capacity, the estimated traffic matrix and the weights of the other links. This can be achieved with standard procedures in OSPF-based environments (see Sec.~\ref{sec:scenario}).

\subsubsection{State Space and Message Passing} 
At each time-step~$t$ of an episode, each link-based agent $v_e \in \mathcal{V},$ feeds its MPNN module with its input features $x_e^t$ to generate its respective initial hidden state $h_e^0$ (Figure \ref{fig:message_passing}.a). In particular, agents consider as input features the current weight $w_e^t$ and the utilization $u_e^t$ $[0,1]$ of the link, and construct their initial link hidden representations $h_e^0$ as a fixed-size vector where the first two components are the input features and the rest is zero-padded. Note that the link utilization can be easily computed by the agent with the information of the estimated traffic matrix and the global link weights locally maintained. Then, the algorithm performs \textit{K} message-passing steps (Figures \ref{fig:message_passing}.b and \ref{fig:message_passing}.c). At each step $k$, the algorithm is executed in a distributed fashion over all the links of the network. 
Particularly, each link-based agent $e \in \mathcal{E}$ receives the hidden states of its neighboring agents $\mathcal{B}(e)$, and combines them individually with its own state $h_e^k$ through the $message$ function (a fully-connected NN).
Then, all these outputs are gathered according to the $aggregation$ function --in our case an element-wise min and max operations-- producing the combination $M_e^k$. Afterwards, another fully-connected NN is used as the $update$ function, which combines the link's hidden state $h_e^k$ with the new aggregated information $M_e^k$, and produces a new hidden state representation for that link ($h_e^{k+1}$). As mentioned above, this process is repeated \textit{K} times, leading to some final link hidden state representations $h_e^K$.

\subsubsection{Action Space} 
In our implementation, each agent $e\in \mathcal{E}$ can only take a single action: to increase its link weight $w_e$ in one unit. In particular, the agent's action selection (Figure \ref{fig:message_passing}.d) is done as follows: first, every agent applies a local readout function --implemented with a fully-connected NN-- to its final hidden state $h_e^K$, from which it obtains the global logit estimate of choosing its action (i.e., increase its link weight) over the actions of the other agents. Then, as previously described in Section \ref{subsec:framework}, these logits are shared among agents in the network, so that each of them can construct the global policy distribution $\pi_\theta$. By sharing the same probabilistic seed, all agents sample locally the same set of actions $A_t$. Finally, agents whose action has been selected increase by one unit the weight of their associated link in its internal global state copy, which is then used to compute the new link utilization $u_e^{t+1}$ under the new weight setting, as well as to initialize its hidden state representation in the next time-step $t+1$.

\subsubsection{Reward Function}
During training, a reward function is computed at each step $t$ of the optimization episode. In our case, given our optimization goal we directly define the reward $r_t$ as the difference of the global maximum link utilization between steps $t$ and $t+1$. Note that this reward can be computed locally at each agent from its global state copy, which is incrementally updated with the new actions applied at each time-step.

\subsection{Training Details} \label{subsec:Training}

The training procedure highly depends on the type of RL algorithm chosen. In our particular implementation, given that we considered an Actor-Critic method (PPO), the objective at training is to optimize the parameters $\{\theta, \phi\}$ so that:
\begin{itemize}
    \item the previously described GNN-based actor $\pi_\theta$ becomes a good estimator of the optimal global policy;
    \item the critic $V_\phi$ learns to approximate the state value function of any global state.
\end{itemize}
As commented in Section \ref{subsec:back-RL}, the goal of the critic is to guide the learning process of the actor; it is no longer needed at execution time. Therefore, taking $V_\phi$ a centralized design would have no impact on the distributed nature of MAGNNETO.

In fact, following the standard approach of MARL systems~\cite{oliehoek2008}, the training of MAGNNETO is performed in a centralized fashion, and such centrality precisely comes from the critic's model. In particular, we have implemented $V_\phi$ as another link-based MPNN, similar to the actor but with a centralized readout that takes as inputs all link hidden states in and outputs the value function estimate. We also considered a MPNN-based critic to exploit the relational reasoning provided by GNNs; however, note that any other alternative design might be valid as well.
    
At a high level, the training pipeline is as follows. First, an episode of length $T$ is generated by following the current policy $\pi_\theta$, while at the same time the critic's value function $V_\phi$ evaluates each visited global state; this defines a trajectory $\{s_t,a_t,r_t,p_t,V_t,s_{t+1}\}_{t=0}^{T-1}$, where $p_t = \pi_\theta(a_t | s_t)$ and $V_t := V_\phi (s_t)$. When the episode ends, this trajectory is used to update the model parameters --through several epochs of minibatch Stochastic Gradient Descent-- by maximizing the global PPO objective $L^{PPO}(\theta, \phi)$ described in \cite{schulman2017proximal}. The same process of generating episodes and updating the model is repeated for a fixed number of iterations to guarantee convergence.

\section{Evaluation} \label{sec:results}

In this section we extensively evaluate MAGNNETO in an intradomain TE scenario: we benchmark it against a curated set of advanced TE optimizers in more than 75 different real-world topologies, using realistic traffic loads. As shown in our experimental results, MAGNNETO achieves similar performance to state-of-the-art TE optimizers with a significantly lower execution time. 
We begin by describing the considered baselines as well as the setup used in our evaluations. The rest of the section is devoted to analyze the results.

\subsection{Baselines} \label{subsec:baselines}

In this section we describe the set of baselines we use to benchmark MAGNNETO in our evaluation. We particularly consider a well-established standard TE mechanism and three advanced TE optimizers.

The first baseline is labeled as "Default OSPF", a simple heuristic widely used in today's ISP networks. In Default OSPF, link weights are inversely proportional to their capacities and traffic is split over multiple paths using Equal-Cost Multi-Path (ECMP). In our experiments, all performance results are expressed in terms of their improvement with respect to Default OSPF.

As state-of-the-art TE benchmarks, we consider the following set of centralized algorithms provided by REPETITA~\cite{gay2017repetita}:
\begin{itemize}
	\item TabuIGPWO (IGP Weight Optimizer, based on \cite{fortz2000internet}): This algorithm runs a Local Search to find the OSPF weights that minimize the load of the maximally-utilized link. TabuIGPWO requires more execution time than the rest of baselines, but represents a classical TE optimizer that operates in the same optimization space than MAGNNETO (i.e., OSPF link weight configuration).
	\item DEFO (Declarative and Expressive Forwarding Optimizer)~\cite{hartert2015declarative}: It uses Constraint Programming and Segment Routing (SR)~\cite{filsfils2015segment} to optimize routing configurations in the order of minutes. To this end, DEFO reroutes traffic paths through a sequence of middlepoints, spreading their traffic over multiple ECMP paths.  
	\item SRLS (Segment Routing and Local Search)~\cite{gay2017expect}: By leveraging Local Search and SR, SRLS achieves similar --or even better-- performance than DEFO at a lower execution time. It also implements ECMP, and reroutes traffic paths through a sequence of middlepoints.
\end{itemize}
Particularly, SRLS and DEFO represent state-of-the-art TE optimizers obtaining close-to-optimal performance on several network optimization goals --one of them being our intradomain TE goal of minimizing the most loaded link. To this end, both optimizers leverage SR, which enables to define overlay paths at a source-destination granularity. In contrast, MAGNNETO and TabuIGPWO operate directly over standard OSPF-based networks with destination-based routing.

\subsection{Experimental Setup}

We compare MAGNNETO against the previously defined TE baselines in all our experimental settings, which involve 82 different real-world topologies: NSFNet, GBN, and GEANT2 from~\cite{rusek2019unveiling}, and 79 networks from the Internet Topology Zoo dataset~\cite{knight2011internet}. In this section we provide more low-level technical details of MAGNNETO's configuration, required to reproduce the results.

Regarding the length $T$ of the training and evaluation RL-based episodes, it varies depending on the network topology size and the number of simultaneous actions allowed (more details below in Sec.~\ref{subsec:multiagent}). At the beginning of each episode, the link weights are randomly selected as an integer in the range $[1,4]$, so our system is evaluated over a wide variety of scenarios with random routing initializations. From that point on, at each step of an episode one or several agents can modify their weight by increasing it by one unit.

Taking~\cite{andrychowicz2021what} as a reference for defining the hyperparameters' values of the solution, we ran several grid searches to appropriately fine-tune the model. The implemented optimizer is Adam with a learning rate of $3$$\cdot$$10^{-4}$, $\beta$=$0.9$, and $\epsilon$=$0.01$. Regarding the PPO setting, the number of epochs for each training episode is set to $3$ with batches of $25$ samples, the discount factor $\gamma$ is set to $0.97$, and the clipping parameter is $0.2$. We implement the Generalized Advantage Estimate (GAE), to estimate the advantage function with $\lambda$=$0.9$. In addition, we multiply the critic loss by a factor of $0.5$, and we implement an entropy loss weighted by a factor of $0.001$. Finally, links' hidden states $h_e$ are encoded as 16-element vectors, and in each MPNN forward propagation $K$=$4$ message passing steps are executed.

For each experiment, we generate two sets of simulated traffic matrices: uniform distribution across source-destination traffic demands, and traffic distributions following a gravity model~\cite{roughan2005simplifying} --which produces realistic Internet traffic patterns. The training process of MAGNNETO highly depends on the topology size; in a machine with a single CPU of 2.20 GHz, it can take from few hours ($\approx$$20$ nodes) to few days (100+ nodes).

\subsection{Multiple Actions and Episode Length} \label{subsec:multiagent}

\begin{figure}[!t]
\centering
    \includegraphics[width=\columnwidth]{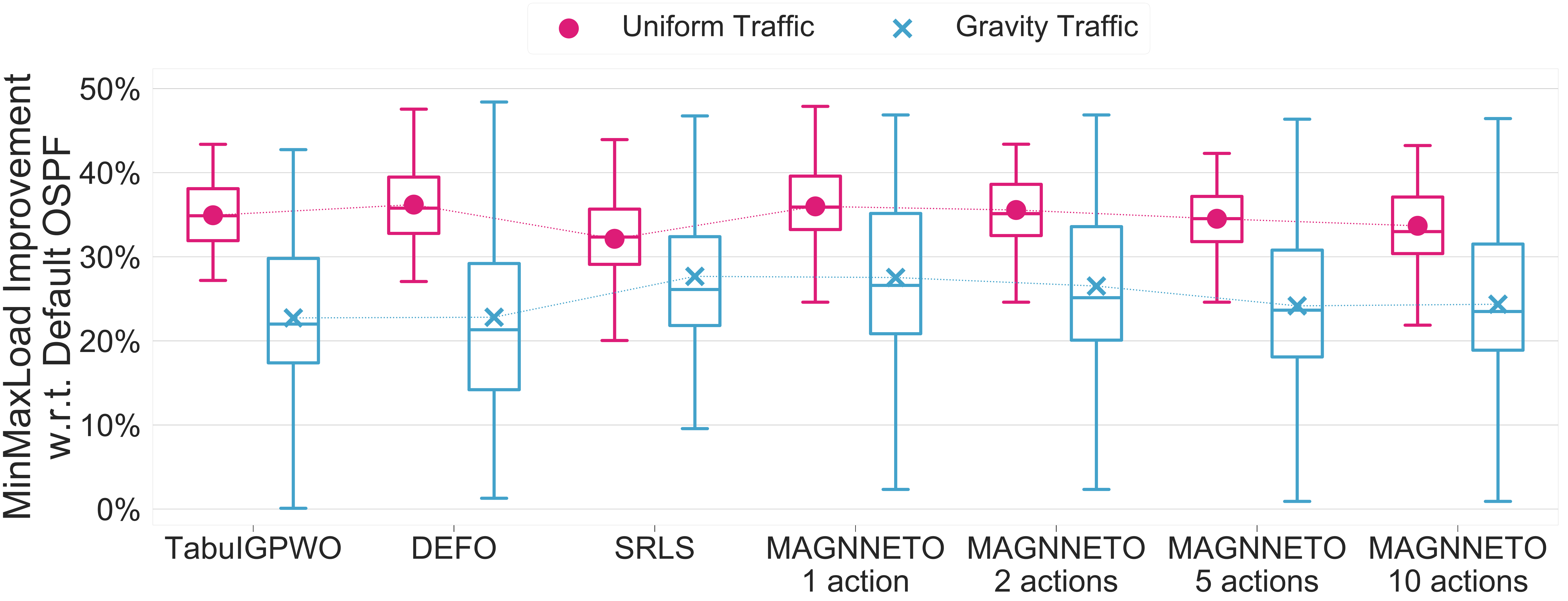}
  \caption{Evaluation of MAGNNETO for different number of simultaneous actions $n \in \{1,2,5,10\}$, each of them considering an episode length of $T=150/n$. The training only considers samples of NSFNet and GEANT2 topologies, and the evaluation is performed over 100 unseen TMs on the GBN topology. Each MAGNNETO model and baseline optimizer is trained and/or evaluated twice for uniform and gravity-based traffic profiles; markers represent the mean of these results, and we also include the corresponding boxplots.}
  \label{fig:multiagent}
  \vspace*{-0.2cm}
\end{figure} 

\begin{figure*}[!t]
    \begin{subfigure}[]{0.99\columnwidth}
	\includegraphics[width=1.0\linewidth]{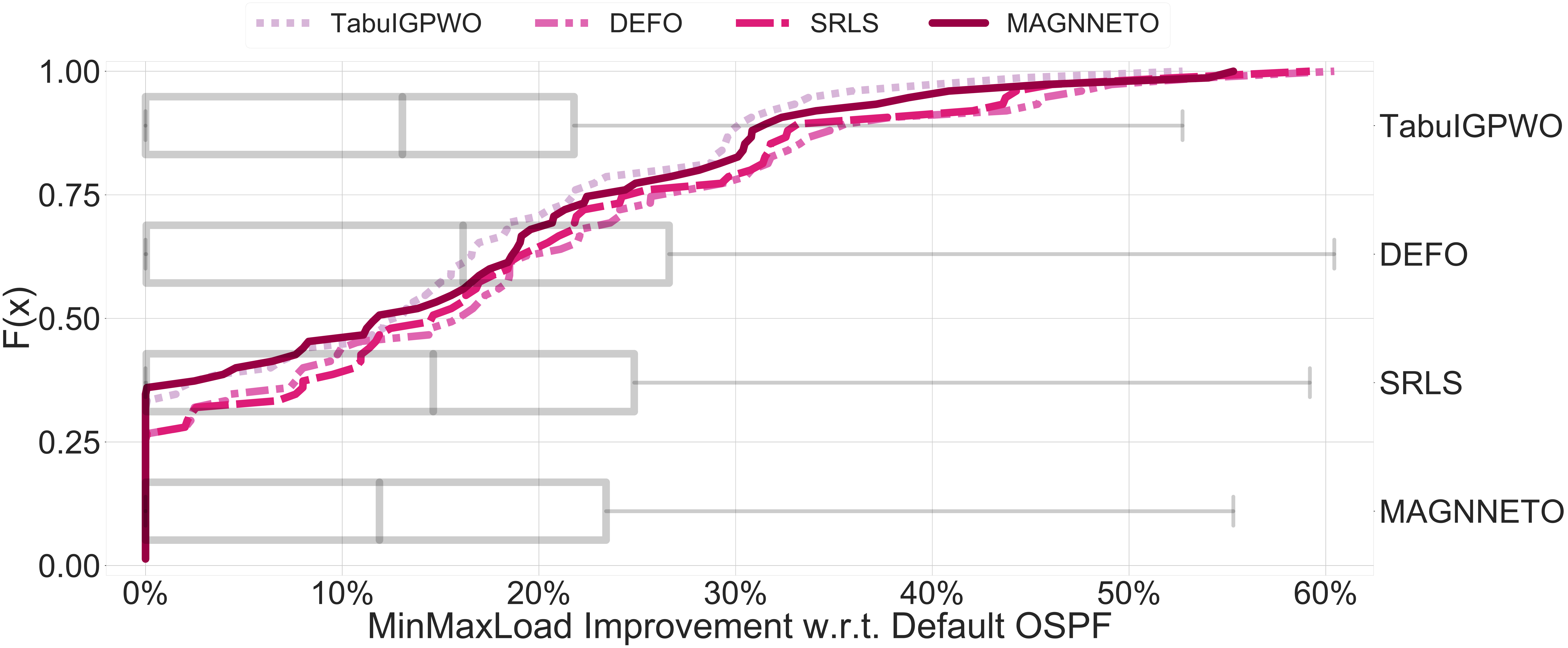}
	\vspace{-0.5cm}
        \caption{TopologyZoo Uniform}
	\label{subfig:topologyzoo_uniform}
    \end{subfigure}
    \begin{subfigure}[]{0.99\columnwidth}
	\includegraphics[width=1.0\linewidth]{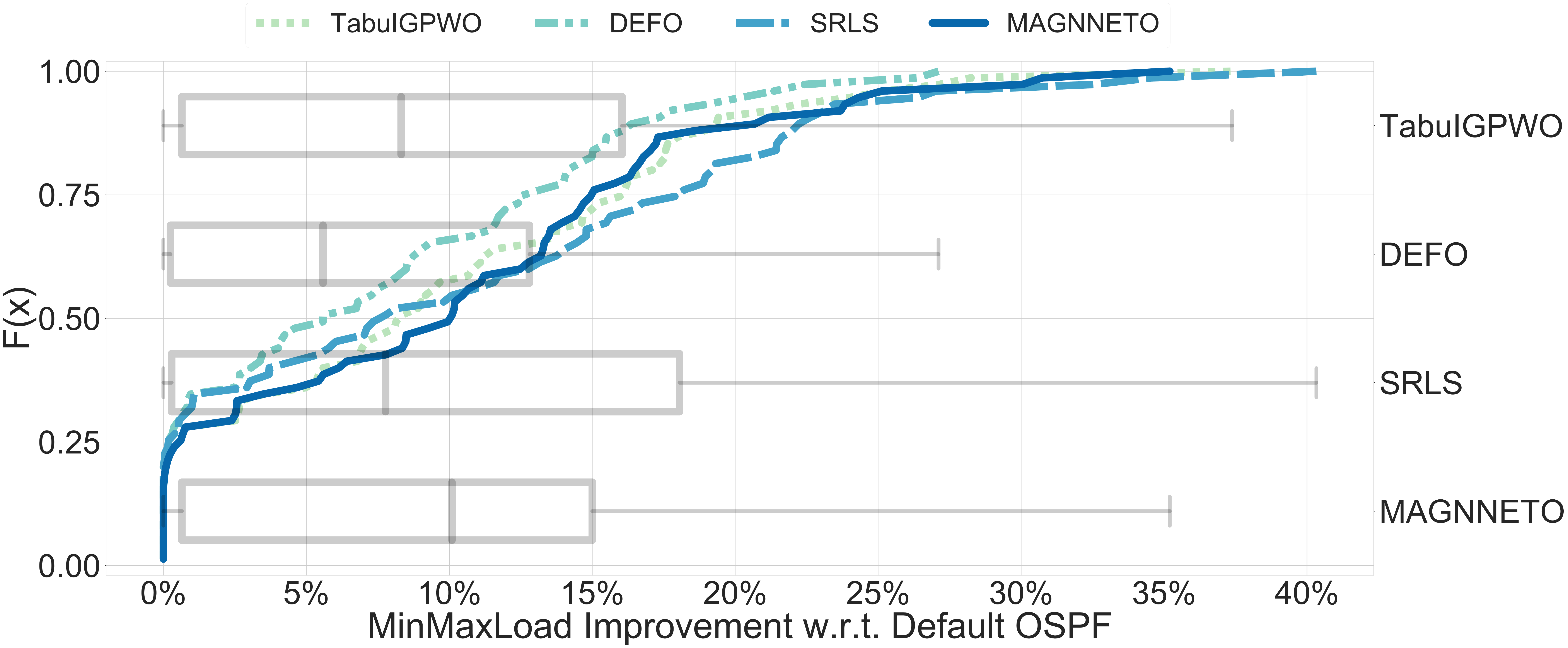}
	\vspace{-0.5cm}
        \caption{TopologyZoo Gravity}
	\label{subfig:topologyzoo_gravity}
    \end{subfigure}
     \caption{Evaluation of MAGNNETO's generalization capability for (a) uniform and (b) gravity traffic. Each point of the CDF corresponds to the mean \textit{MinMaxLoad} improvement over 100 TMs for one of the 75 evaluation topologies from Topology Zoo\cite{knight2011internet}, and boxplots are computed based on these mean improvement values as well. Both the uniform (a) and gravity (b) MAGNNETO models evaluated were trained exclusively on samples from the NSFNet and GEANT2 topologies~\cite{rusek2019unveiling}.}
  \label{fig:topologyzoo}
\end{figure*}

As previously mentioned in Section~\ref{sec:architecture}, there is a relevant hyperparameter that needs to be further addressed: the episode length $T$ of RL-based episodes, which represents the maximum number of optimization steps that MAGNNETO needs to execute before producing a good set of link weights. In this section we provide more details about its definition in terms of the topology size and the number of simultaneous actions.

Let $n$ be such maximum number of simultaneous actions allowed at each time-step $t$ of the episode. When imposing $n$$=$$1$ --i.e., only one link weight changes per time-step--,  we have empirically found that MAGNNETO requires an episode length of $\approx$$2$$-$$3$ times the number of links in the network to reach its best performance. This is in line to what we already observed in our preliminary work~\cite{icnp-paper}. However, whereas \cite{icnp-paper} was subject to $n$$=$$1$ by design, MAGNNETO allows taking $n$$>$$1$ actions at each time-step, which can potentially reduce the number of required optimization steps (i.e., speed up the optimization process).

Figure~\ref{fig:multiagent} shows that the length $T$ of the episode \mbox{--which} directly relates to the execution time-- can be reduced proportionally by $n$ without a noticeable performance loss. In particular, the model with $n$$=$$10$ actually reduces by one order of magnitude the execution time of the 1-action model, but still achieves comparable performance to the state-of-the-art optimizers of our benchmark --for both traffic profiles, and evaluating on a topology not previously seen in training.

Given the good trade-off that provides allowing more than one action at each time-step, for the rest of our experiments we fine-tuned the number of actions $n$ and the episode length $T$ to balance a competitive performance with the minimum possible execution time. Later in Section \ref{subsec:res-cost} we will analyze in detail the execution cost of MAGNNETO.

\subsection{Generalization over Unseen Topologies}

In Section~\ref{sec:introduction} we argued the importance of generalization in ML-based solutions, which refers to the capability of the solution to operate successfully in other networks where it was not trained. In this section, we bring MAGNNETO under an intensive evaluation in this regard. 

In our experiments, MAGNNETO only observes NSFNet (14 nodes, 42 links) and GEANT2 (24 nodes, 74 links) samples during training~\cite{rusek2019unveiling}, whereas the evaluation is performed over a subset of 75 networks from the Topology Zoo dataset~\cite{knight2011internet} including topologies ranging from 11 to 30 nodes, and from 30 to 90 links. More in detail: 
\begin{itemize}
    \item We train two MAGNNETO models, one for each traffic profile (uniform and gravity).
    \item Each model is trained observing 50 different TMs \mbox{--either} uniform or gravity-based, depending on the \mbox{model--} alternating between the NSFNet and GEANT2 topologies.
    \item Each of these two trained models is evaluated over 100 different TMs --again, either uniform or gravity-based-- on each of the 75 topologies from Topology Zoo.
\end{itemize}

Overall, this experimental setup comprises $7,500$ evaluation runs for each traffic profile, which we summarize in Figures \ref{subfig:topologyzoo_uniform} and \ref{subfig:topologyzoo_gravity}, respectively for uniform and gravity-based loads. In particular, note that we first compute the mean \textit{MinMaxLoad} improvement of MAGNNETO --and the baselines-- over the 100 TMs of each evaluation network, obtaining a single value for each of the 75 topologies. Thus, in these figures we represent the corresponding CDF and boxplot of the 75-sized vector of mean improvement values for each TE optimizer.

In both traffic scenarios MAGNNETO achieves comparable performance to the corresponding best performing benchmark --DEFO when considering uniform traffic and SRLS for gravity. In fact, MAGNNETO outperforms TabuIGPWO, improves DEFO with gravity-based traffic, and lies within a $2\%$ average improvement difference with respect to SRLS in both cases. We reckon that these represent remarkable results on generalization; as far as we know, this is the first time that a ML-based model consistently obtains close performance to state-of-the-art TE optimizers on such a large and diverse set of topologies not previously seen in training.

\begin{figure*}[!t]
    \begin{subfigure}[]{0.5\columnwidth}
	\includegraphics[width=1.0\linewidth]{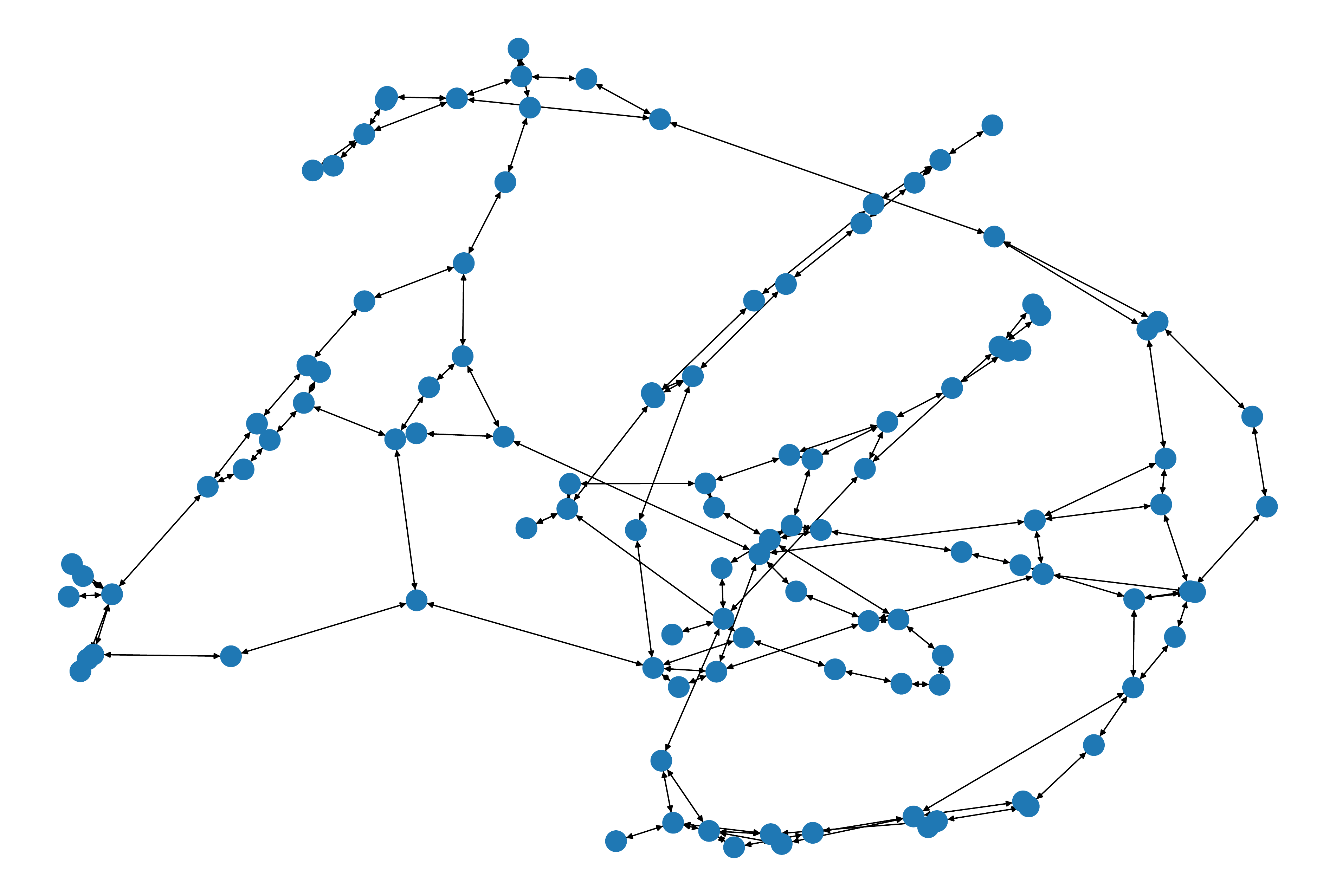}
	\label{subfig:raw_NSFNet_uniform}
	\vspace{-0.5cm}
        \caption*{\textsc{\textbf{I.} Interoute}}
	\vspace{0.3cm}
    \end{subfigure}
    \begin{subfigure}[]{0.5\columnwidth}
	\includegraphics[width=1.0\linewidth]{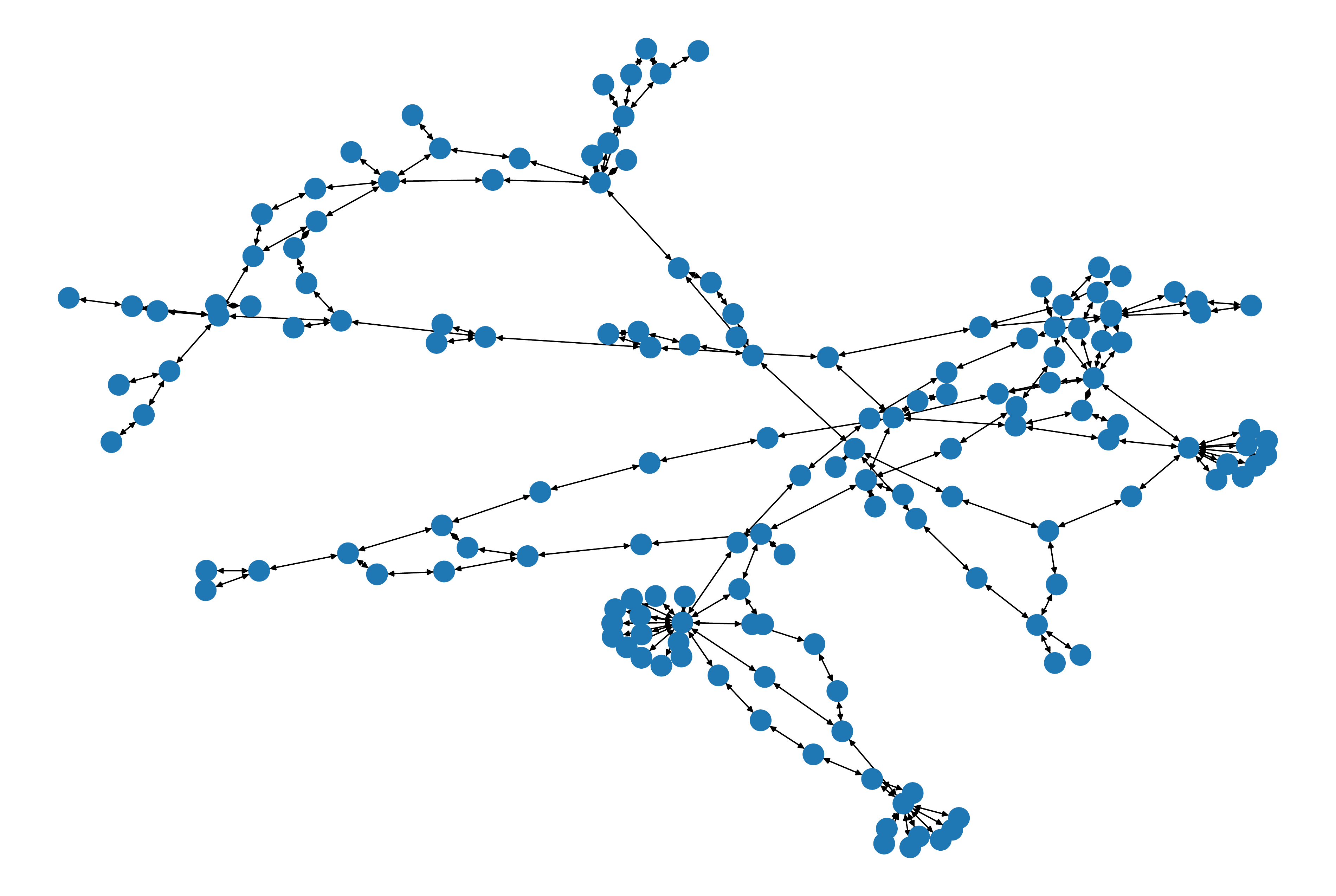}
	\label{subfig:raw_NSFNet_uniform}
	\vspace{-0.5cm}
        \caption*{\textsc{\textbf{II.} Colt}}
	\vspace{0.3cm}
    \end{subfigure}
    \begin{subfigure}[]{0.5\columnwidth}
	\includegraphics[width=1.0\linewidth]{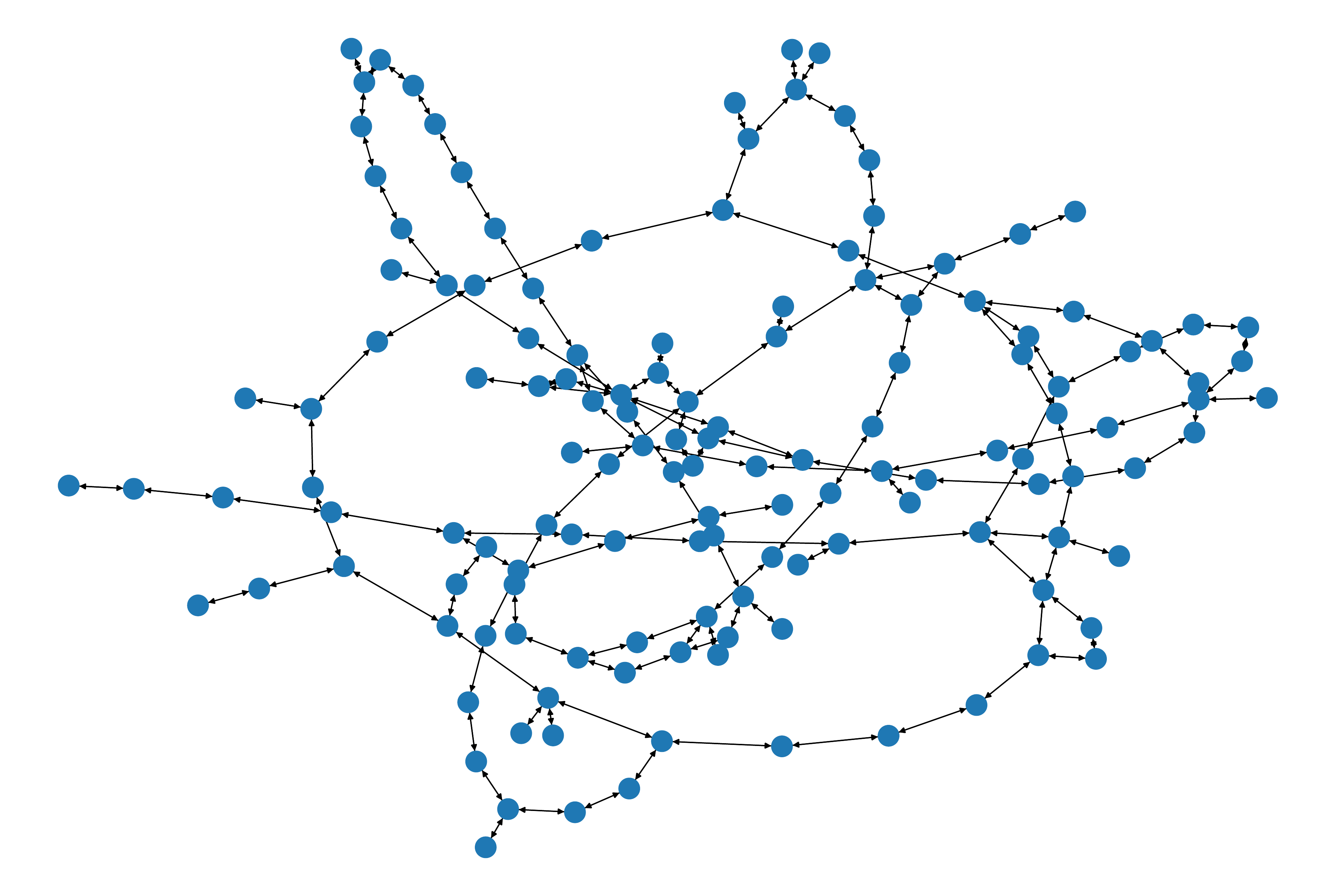}
	\label{subfig:raw_NSFNet_uniform}
	\vspace{-0.5cm}
        \caption*{\textsc{\textbf{III.} DialtelecomCz}}
	\vspace{0.3cm}
    \end{subfigure}
    \begin{subfigure}[]{0.5\columnwidth}
	\includegraphics[width=1.0\linewidth]{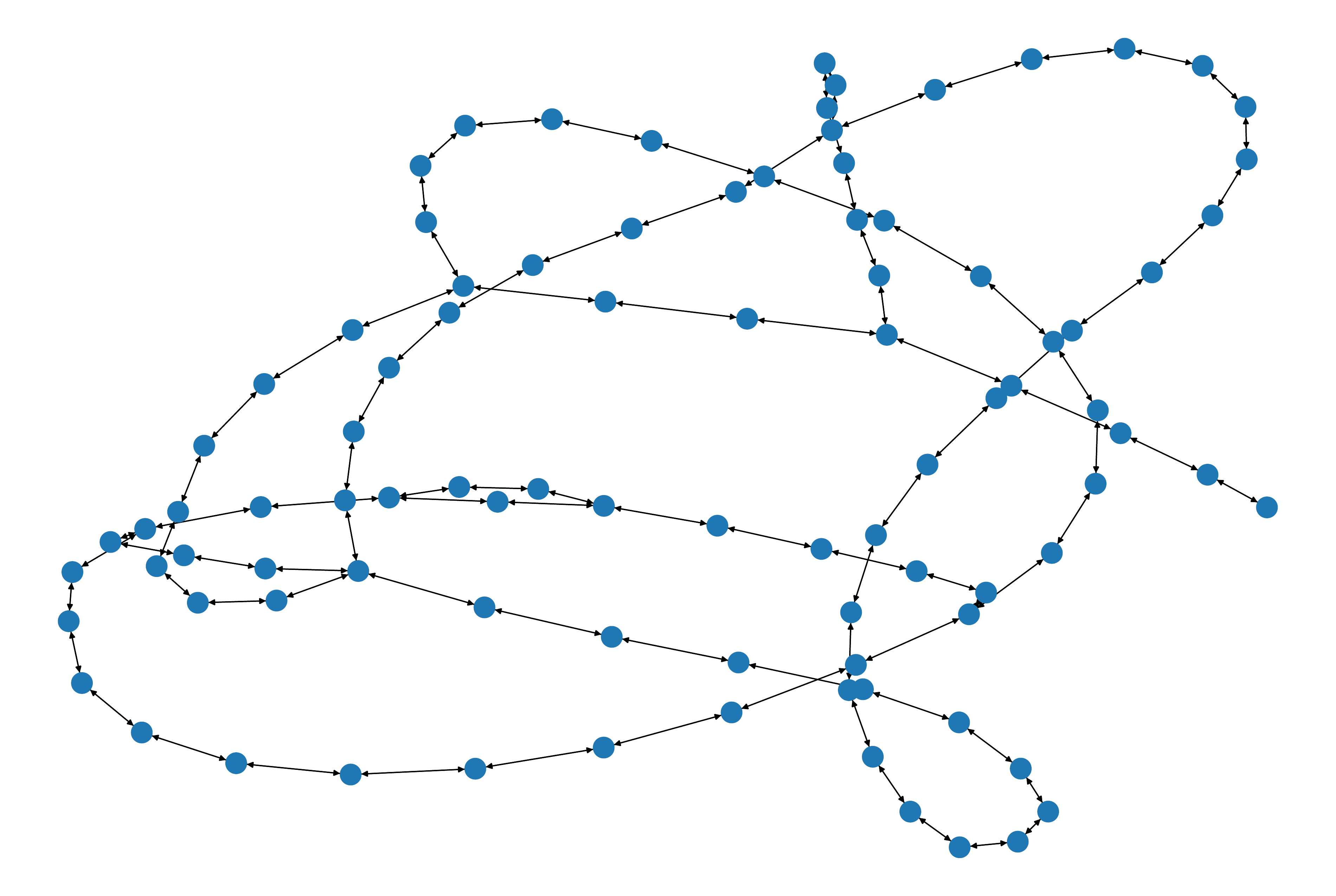}
	\label{subfig:raw_NSFNet_uniform}
	\vspace{-0.5cm}
        \caption*{\textsc{\textbf{IV.} VtlWavenet2011}}
	\vspace{0.3cm}
    \end{subfigure}
    \begin{subfigure}[]{0.5\columnwidth}
	\includegraphics[width=1.0\linewidth]{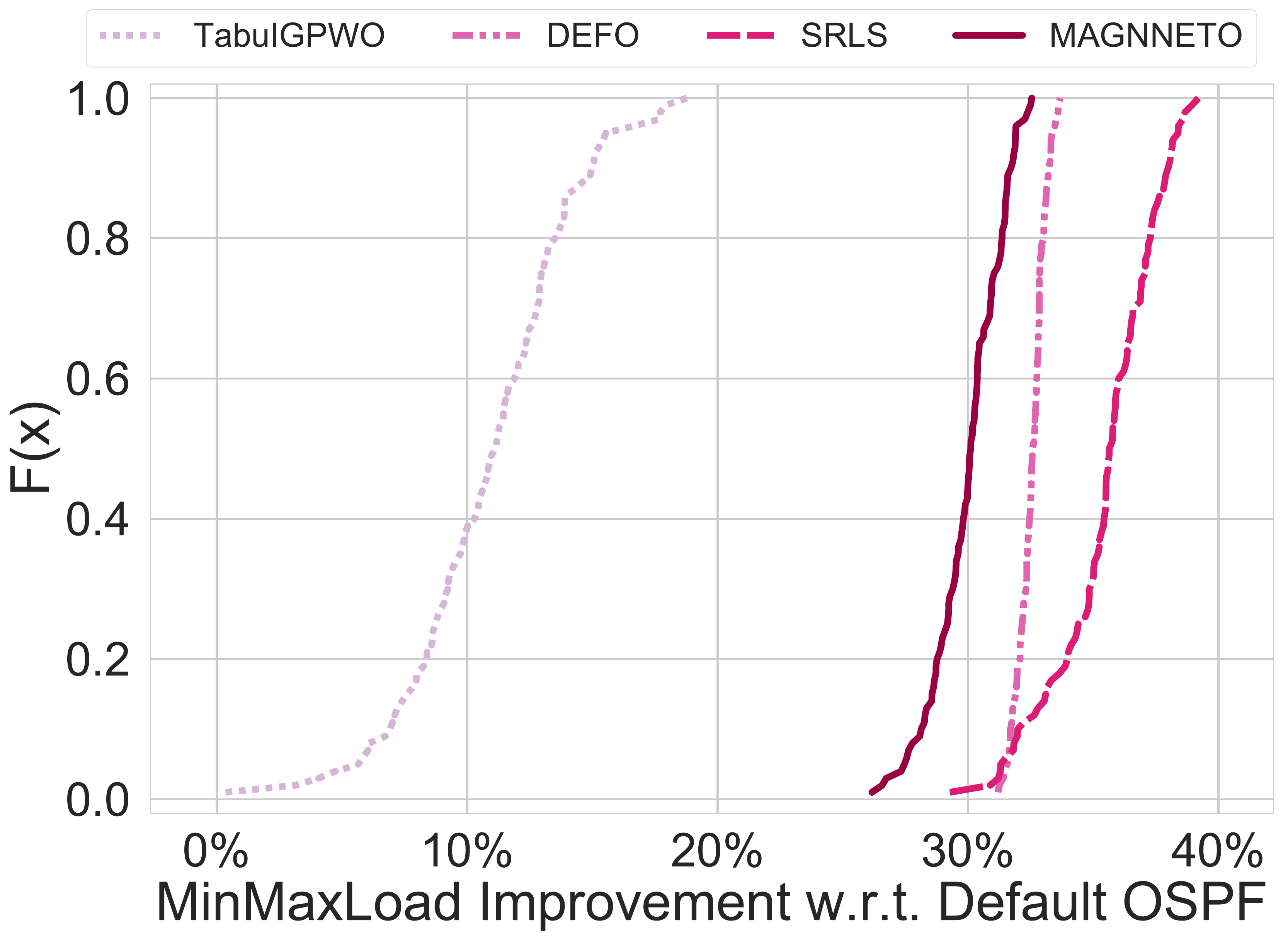}
	\label{subfig:raw_NSFNet_uniform}
	\vspace{-0.5cm}
        \caption{Interoute Uniform}
	\vspace{0.3cm}
    \end{subfigure}
    \begin{subfigure}[]{0.5\columnwidth}
	\includegraphics[width=1.0\linewidth]{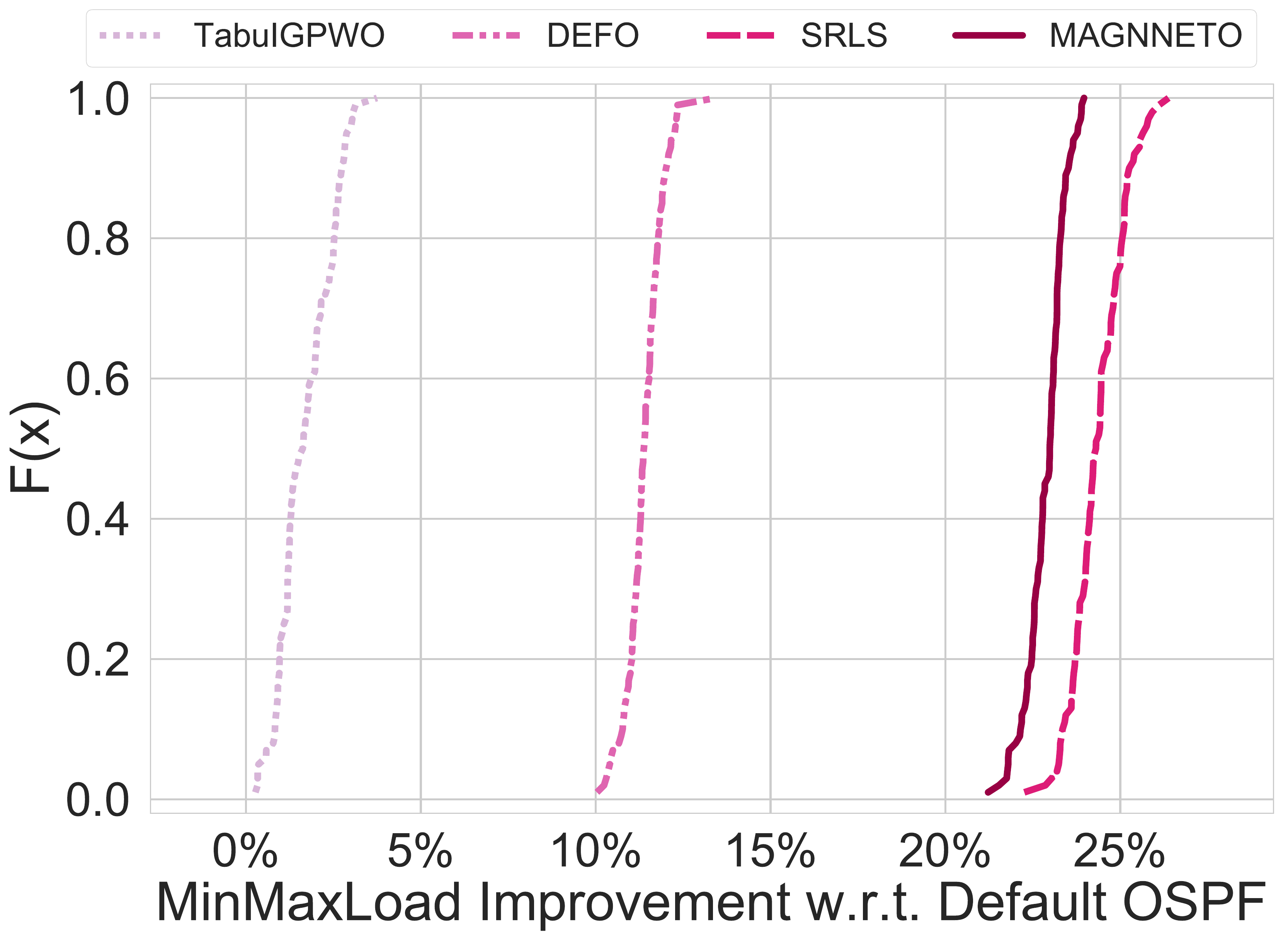}
	\label{subfig:raw_NSFNet_uniform}
	\vspace{-0.5cm}
        \caption{Colt Uniform}
	\vspace{0.3cm}
    \end{subfigure}
    \begin{subfigure}[]{0.5\columnwidth}
	\includegraphics[width=1.0\linewidth]{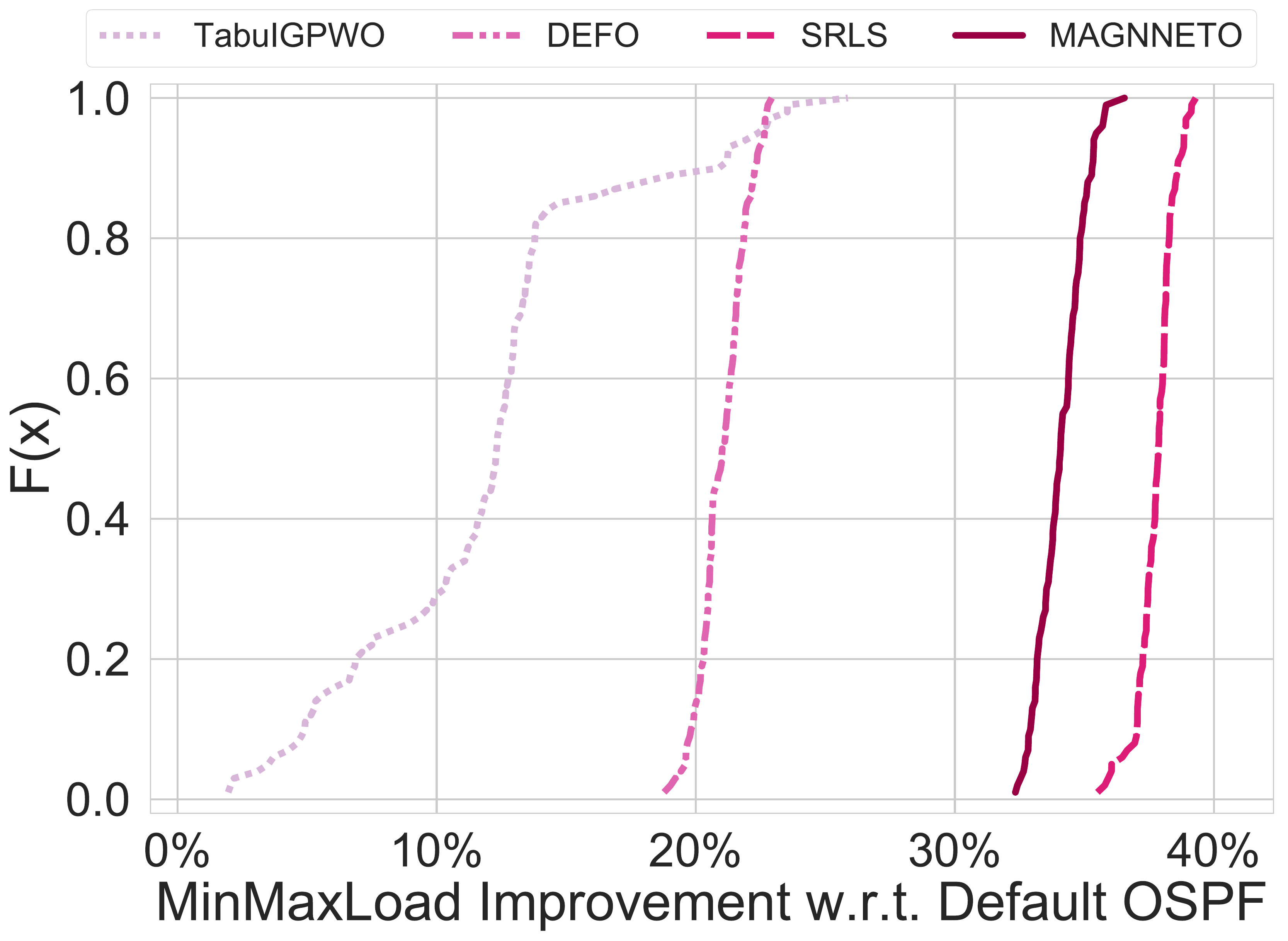}
	\label{subfig:raw_NSFNet_uniform}
	\vspace{-0.5cm}
        \caption{DialtelecomCz Uniform}
	\vspace{0.3cm}
    \end{subfigure}
    \begin{subfigure}[]{0.5\columnwidth}
	\includegraphics[width=1.0\linewidth]{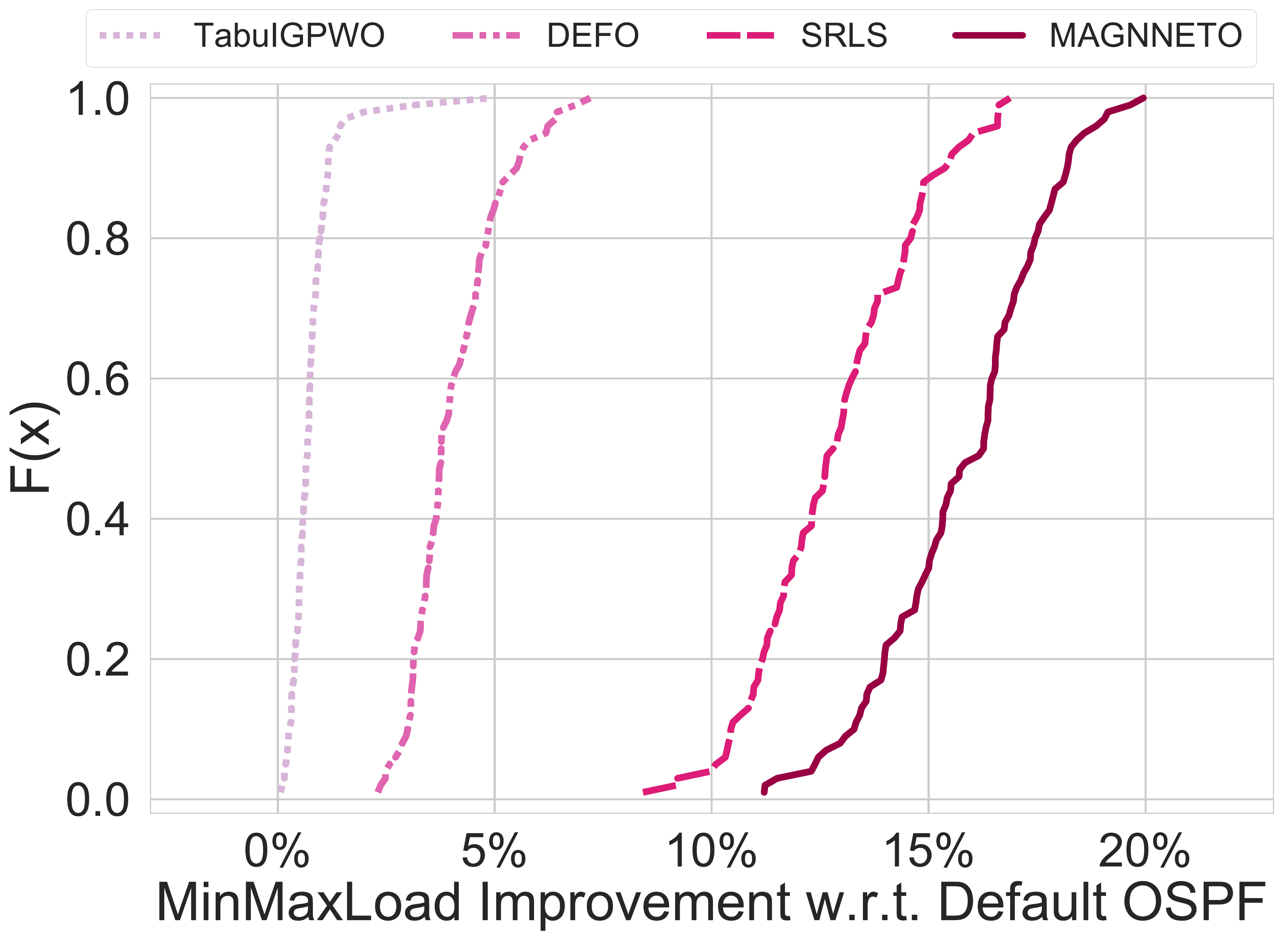}
	\label{subfig:raw_NSFNet_uniform}
	\vspace{-0.5cm}
        \caption{VtlWavenet2011 Uniform}
	\vspace{0.3cm}
    \end{subfigure}
    \begin{subfigure}[]{0.5\columnwidth}
	\includegraphics[width=1.0\linewidth]{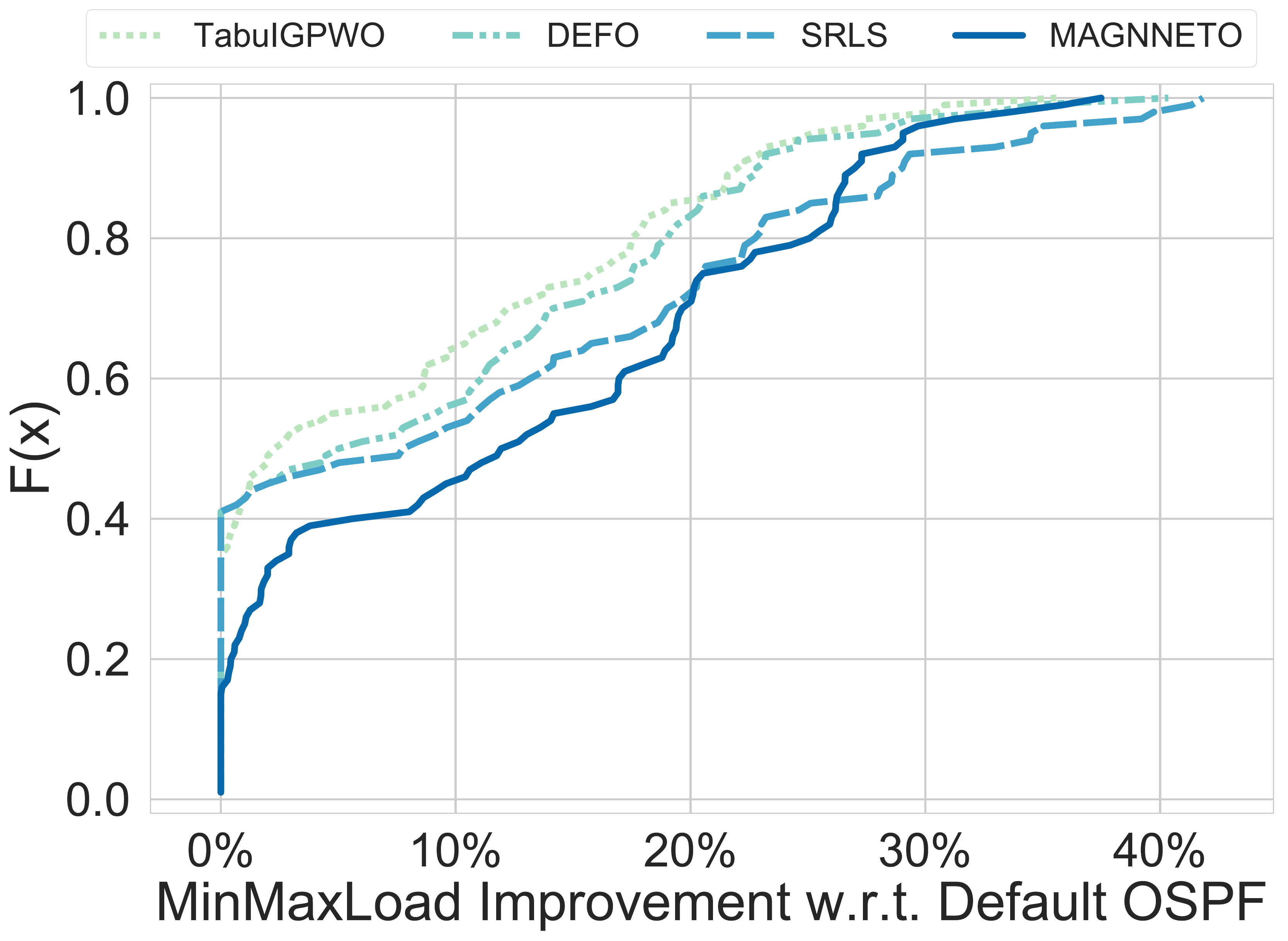}
	\label{subfig:raw_GEANT2_uniform}
	\vspace{-0.5cm}
        \caption{Interoute Gravity}
	\vspace{0.3cm}
    \end{subfigure}
    \begin{subfigure}[]{0.5\columnwidth}
	\includegraphics[width=1.0\linewidth]{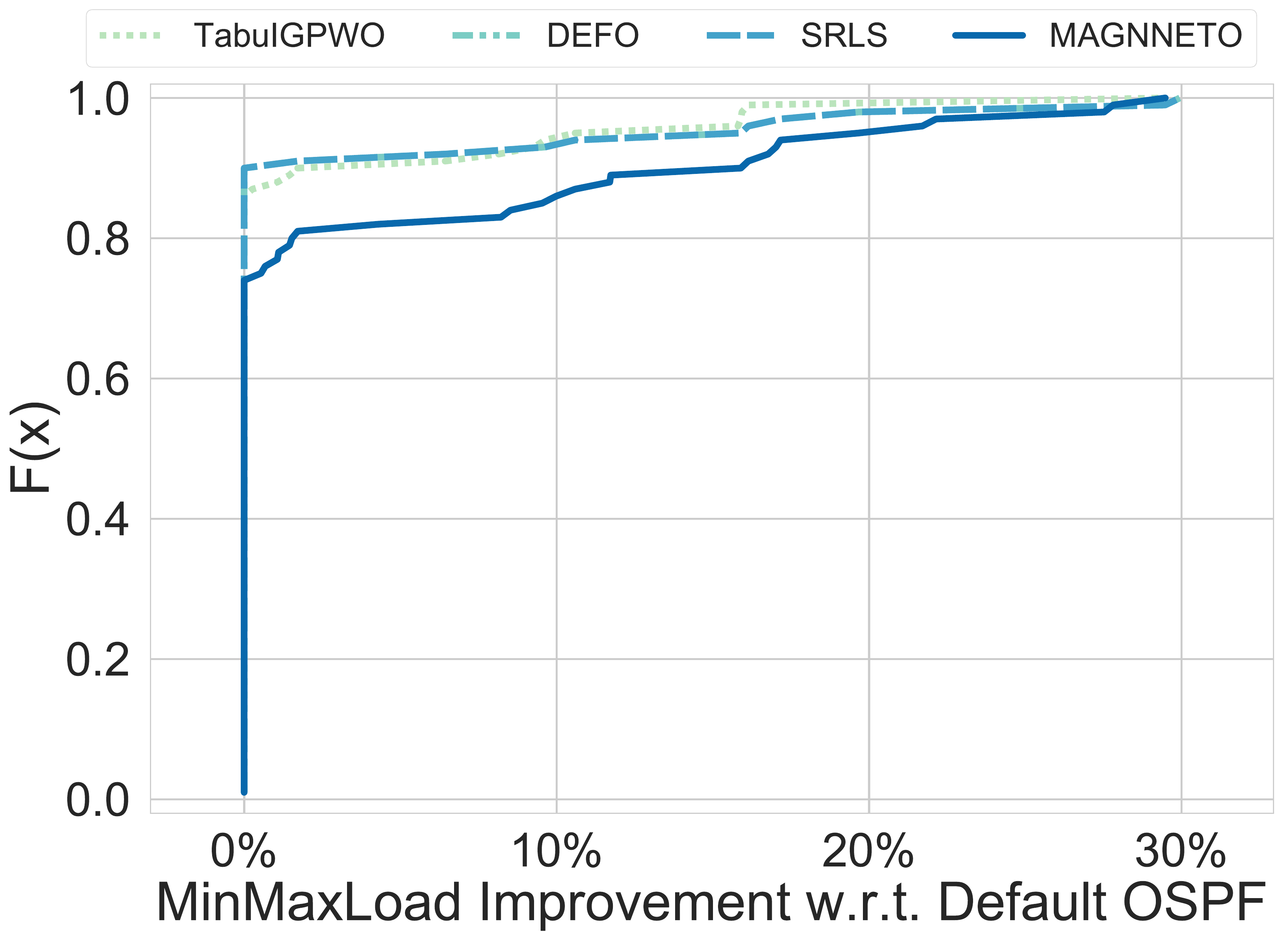}
	\label{subfig:raw_GEANT2_uniform}
	\vspace{-0.5cm}
        \caption{Colt Gravity}
	\vspace{0.3cm}
    \end{subfigure}
    \begin{subfigure}[]{0.5\columnwidth}
	\includegraphics[width=1.0\linewidth]{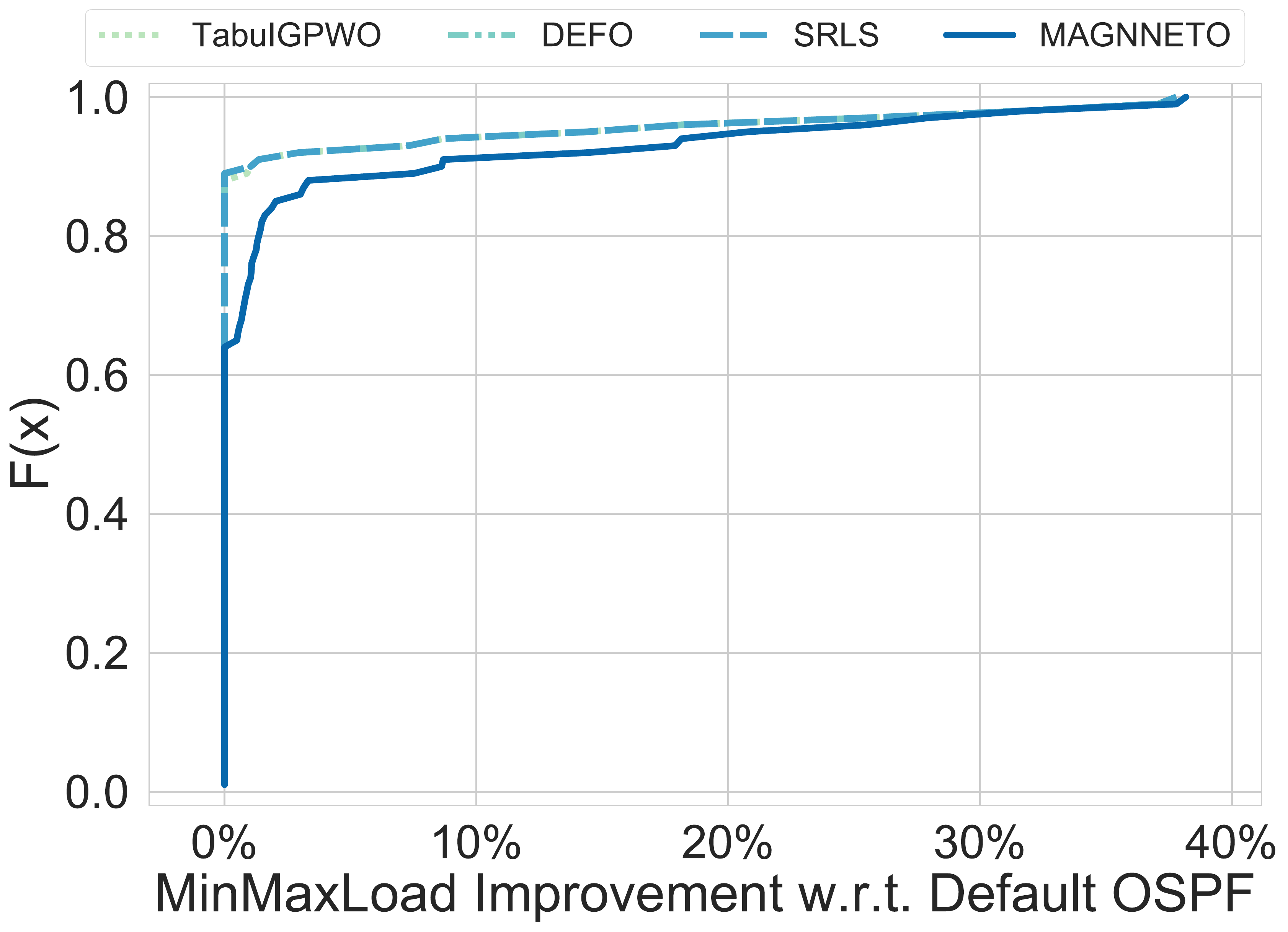}
	\label{subfig:raw_GEANT2_uniform}
	\vspace{-0.5cm}
        \caption{DialtelecomCz Gravity}
	\vspace{0.3cm}
    \end{subfigure}
    \begin{subfigure}[]{0.5\columnwidth}
	\includegraphics[width=1.0\linewidth]{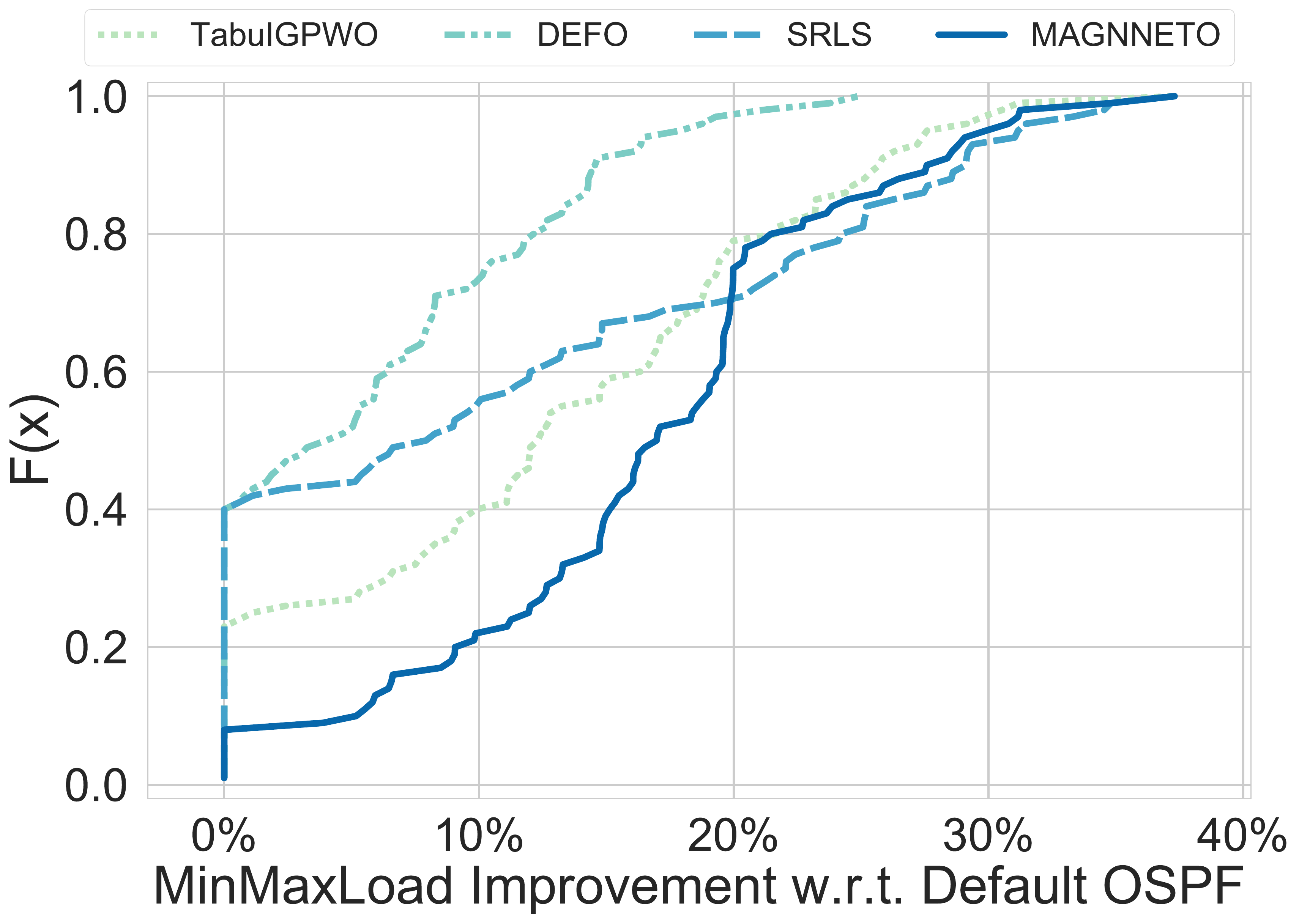}
	\label{subfig:raw_GEANT2_uniform}
	\vspace{-0.5cm}
        \caption{VtlWavenet2011 Gravity}
	\vspace{0.3cm}
    \end{subfigure}
    \vspace{-0.25cm}
     \caption{Evaluation of MAGNNETO on traffic changes in four large real-world topologies (I-IV) from the Topology Zoo dataset~\cite{knight2011internet}, both for uniform ((a)-(d)) and gravity-based ((e)-(h)) traffic loads. A MAGNNETO model is trained for each network and traffic profile, and then evaluated on the same topology over 100 unseen TMs. CDFs represent the \textit{MinMaxLoad} improvement results of each optimizer for those 100 evaluation TMs.}
  \label{fig:large}
\end{figure*}

\subsection{Traffic Changes in Large Topologies}

After evaluating the generalization capabilities of MAGNNETO, we aim to test the performance of our method over traffic changes in large networks, where the combinatorial of the optimization process might dramatically increase. Having considered networks up to 30 nodes and 90 links so far, for this set of experiments we arbitrarily select four large real-world topologies from Topology Zoo~\cite{knight2011internet}: Interoute (110 nodes, 294 links), Colt (153 nodes, 354 links), DialtelecomCz (138 links, 302 links) and VtlWavenet2011 (92 nodes, 192 links). Figures \ref{fig:large}.I-IV depict these topologies.

In these experiments, for each traffic profile (uniform or gravity) we train a MAGNNETO model on each network. Then, we evaluate models on the same topology where they were trained, over 100 different TMs not previously seen in training. Figures \ref{fig:large}.a-d and  \ref{fig:large}.e-f show the corresponding CDF of all these evaluations, considering uniform and gravity traffic loads respectively.

As we can see, with uniform traffic SRLS is clearly the best performing baseline, achieving a remarkable overall improvement gap with respect to the other two benchmarked optimizers. However, in this scenario MAGNNETO is able to obtain similar improvements to SRLS, slightly outperforming it in VtlWavenet2011. On the other hand, results with gravity-based traffic suggest that Default OSPF already provides with low-congested routing configurations in scale-free networks when considering more realistic traffic. Despite this fact, MAGNNETO turns out to be the overall winner in the comparison with gravity loads, consistently achieving lower congestion ratios for a large number of TMs in all four topologies.

In short, in all scenarios MAGNNETO attained equivalent --or even better-- performance than the advanced TE optimizers benchmarked. These results evince its potential to successfully operate in large computer networks.

\subsection{Execution Cost} \label{subsec:res-cost}

\begin{table*}[ht]
\centering
\begin{tabular}{@{}lccccccc@{}}
\toprule
                             & \textbf{NSFNet} \hspace*{1.5mm}  &   \textbf{GBN}   & \textbf{GEANT2}  &  \textbf{VtlWavenet2011}    & \textbf{Interoute}  & \textbf{DialtelecomCz} & \textbf{Colt}   \\ \midrule
(\#nodes, \#links)               &     (14,42)          &    (17,52)           &      (24,74)         &       (92,192)       &     (110,294)    &     (138,302)    &     (153,354)         \\
MAGNNETO Link Overhead$^{*}$ (MB/s) &     1.20         &     1.32         &     1.20         &       0.83       &     1.28     &     0.91    &     1.01         \\
Execution Time (s)          &             &   &            &             &       &        &            \\
\quad TabuIGPWO~\cite{fortz2000internet}         & \hspace*{0.3cm} 600 \hspace*{0.3cm}  & \hspace*{0.3cm} 600 \hspace*{0.3cm}  & \hspace*{0.3cm} 600 \hspace*{0.3cm}  & \hspace*{0.3cm} 600 \hspace*{0.3cm}  & \hspace*{0.3cm} 600 \hspace*{0.3cm}  & \hspace*{0.3cm} 600 \hspace*{0.3cm}  & \hspace*{0.3cm} 600 \hspace*{0.3cm}     \\
\quad DEFO~\cite{hartert2015declarative}          & 180  & 180  & 180  & 180  & 180  & 180  & 180         \\
\quad SRLS~\cite{gay2017expect}          & 60  & 60  & 60  & 60  & 60  & 60  & 60          \\
\quad MAGNNETO [$n$ actions]      &$0.08 / n$&$0.12/ n$&$0.16/ n$& $0.42 / n$     &$0.64 / n$&    $0.66 / n$    &   $0.78 / n$        \\
 \bottomrule
\multicolumn{6}{l}{$^{*}$Average value, with an extra $20\%$ message size for headers and metadata.}
\end{tabular}
\caption{Cost of MAGNNETO: Average link overhead and execution time --in terms of the maximum number of simultaneous actions allowed-- for variable-sized network topologies.}
\label{tab:cost}
\end{table*}

Lastly, in this section we evaluate the execution cost of MAGNNETO. In particular, we measure the impact of the message communications involved when running our distributed solution, as well as compare its execution time against the considered set of state-of-the-art TE baselines; Table~\ref{tab:cost} gathers these results for several variable-sized networks used in the previous evaluations.

Taking into account the recommendations of REPETITA~\cite{gay2017expect}, as well as analyzing the results provided in the original works~\cite{fortz2000internet,hartert2015declarative,gay2017expect}, we defined the following running times for each of our benchmarks: 10 minutes for TabuIGPWO, 3 minutes for DEFO, and 1 minute for SRLS.

At first glance, the execution time of MAGNNETO becomes immediately its most remarkable feature. Particularly, it is able to obtain subsecond times even for the larger network of our evaluation (Colt).
Indeed, as previously discussed in Section~\ref{subsec:multiagent} these times could be further reduced by allowing multiple simultaneous actions. For instance, by considering up to 10 simultaneous actions, MAGNNETO can run 3 orders of magnitude faster than the most rapid state-of-the-art TE optimizer. This relevant difference can be explained by the fact that MAGNNETO's distributed execution naturally parallelizes the global optimization process across all network devices (i.e., routers); in contrast, typical TE optimizers rely on centralized frameworks that cannot benefit from this.

Such decentralization comes at the expense of the extra message overhead generated by the MPNN. In this context, \mbox{Table~\ref{tab:cost}} shows that the link overhead produced by MAGNNETO (few MB/s) can reasonably have a negligible impact in today's real-world networks with 10G/40G (or even more) interfaces. Moreover, note that this cost is quite similar in all topologies; this is as expected, given that the messaging overhead of the GNN-based communications is directly proportional to the average node degree of the network, and computations are distributed among all nodes. 

To sum up, our results show that MAGNNETO is able to attain equivalent performance to state-of-the-art centralized TE optimizers --even in topologies not previously seen in training-- with significantly lower execution time, and with an affordable message communication overhead.

\section{Related Work} \label{sec:related}

Recently, numerous solutions based on Deep Reinforcement Learning (DRL) have been proposed to solve complex networking problems, especially in the context of routing optimization and TE~\cite{valadarsky2017learning,drl-giorgio,suarez2019feature}. However, current state-of-the-art RL-based TE solutions fail to generalize to unseen scenarios (e.g., different network topologies) as the implemented traditional neural networks (e.g., fully connected, convolutional) are not well-suited to learn and generalize over data that is inherently structured as graphs. In \cite{xu2018experience}, the authors design a DRL-based architecture that obtains better results than Shortest Path and Load Balancing routing. 
Regarding MARL-based solutions~\cite{ding2020packet, mao2019neighborhood}, most of them suffer from the same lack of topology generalization. An exception to that is the work of~\cite{geng2020multi}, an interesting MARL approach for multi-region TE that consistently outperforms ECMP in several scenarios, although it is not benchmarked against state-of-the-art TE optimizers.

GNNs~\cite{gori2005new, scarselli2008graph}, and in particular Message Passing Neural Networks (MPNN) \cite{gilmer2017neural}, precisely emerged as specialized methods for dealing with graph-structured data; for the first time, there was an AI-based technology able to provide with topology-aware systems.
In fact, GNNs have recently attracted a large interest in the field of computer networks for addressing the aforementioned generalization limitations. The work from \cite{rusek2019unveiling} proposes to use GNN to predict network metrics and a traditional optimizer to find the routing that minimizes some of these metrics (e.g., average delay). 
Authors of \cite{almasandeep} propose a novel architecture for routing optimization in Optical Transport Networks that embeds a GNN into a centralized, single-agent RL setting that is compared against Load Balancing routing. 

Narrowing down the use case to intradomain TE, we highlight the work of~\cite{geyer2018learning}, whose premise is similar to ours: the generation of easily-scalable, automated distributed protocols. For doing so, the authors also use a GNN, but in contrast to our approach they are focused on learning routing strategies that directly imitate already existing ones --shortest path and min-max routing-- and compare their solution against these ones. This is the reason why they did not implement a RL-based approach, but instead a semi-supervised learning algorithm, therefore guiding the learning process with explicit labeled data. In fact, so far the very few works that combine GNNs with a MARL framework~\cite{jiang2020graph, su2020counterfactual} are theoretical papers from the ML community, and none of them apply to the field of networking.

\section{Conclusions} \label{sec:conclusions}

Intradomain Traffic engineering (TE) is nowadays among the most common network operation tasks, and has a major impact on the performance of today's ISP networks. As such, it has been largely studied, and there are already some well-established TE optimizers that deliver near-optimal performance in large-scale networks. During the last few years, state-of-the-art TE solutions have systematically competed for reducing execution times~(e.g., DEFO~\cite{hartert2015declarative}, SRLS~\cite{gay2017expect}), thus scaling better to carrier-grade networks and achieving faster reaction to traffic changes. In this context, ML has attracted interest as a suitable technology for achieving faster execution of TE tasks and --as a result-- during recent years the networking community has devoted large efforts to develop effective ML-based TE solutions~\cite{valadarsky2017learning,xu2018experience,drl-giorgio,geng2020multi}. However, at the time of this writing no ML-based solution had shown to outperform state-of-the-art TE optimizers.

In this paper we have presented MAGNNETO, a novel ML-based framework for intradomain TE optimization. Our system implements a novel distributed architecture based on Multi-Agent Reinforcement Learning and Graph Neural Networks. In our evaluation, we have compared MAGNNETO with a set of non-ML-based TE optimizers that represent the state of the art in this domain. After applying our system to 75+ real-world topologies, we have observed that it achieves comparable performance to the reference TE benchmarks. However, MAGNNETO offers considerably faster operation than these state-of-the-art TE solutions, reducing execution times from several minutes to sub-second timescales in networks of 100+ nodes. In this context, MAGNNETO was especially designed to perform several actions at each RL optimization step, which enables to considerably accelerate the optimization process. Particularly, we have seen that our system was able to perform up to 10 actions in parallel with no noticeable decrease in performance. These results lay the foundations for a new generation of \mbox{ML-based} systems that can offer the near-optimal performance of traditional TE techniques while reacting much faster to traffic changes.

Last but not least, we have shown that the proposed system offers strong generalization power over networks unseen during the training phase, which is an important characteristic from the perspective of deployability and commercialization. Particularly, generalization enables to train ML-based products in controlled testbeds, and then deploy them in different real-world networks in production. However, this property has been barely addressed by prior ML-based TE solutions. In contrast, MAGNNETO has demonstrated to generalize succesfully over a wide and varied set of 75 real-world topologies unseen during training. The main reason behind this generalization capability is that the proposed system implements internally a GNN that structures and processes network information as graphs, and computes the information on distributed agents that communicate with their neighbors according to the underlying graph structure (i.e., the network topology).

\section*{Acknowledgment}

This publication is part of the Spanish I+D+i project TRAINER-A (ref.~PID2020-118011GB-C21), funded by MCIN/ AEI/10.13039/501100011033. This work is also partially funded by the Catalan Institution for Research and Advanced Studies (ICREA), the Secretariat for Universities and Research of the Ministry of Business and Knowledge of the Government of Catalonia, and the European Social Fund.

\bibliographystyle{ieeetr} 
\bibliography{References.bib}

\end{document}